\documentclass[sigplan,screen]{acmart}
\usepackage[utf8]{inputenc}

\setcopyright{none}
\renewcommand\footnotetextcopyrightpermission[1]{}

\title{Parallel Algorithms for Masked Sparse Matrix-Matrix Products}
\author{Sr\dj{}an Milakovi\'{c}}
\affiliation{%
    \institution{Rice University}
    \city{Houston}
    \state{TX}
    \country{USA}}
    \email{sm108@rice.edu}
\author{Oguz Selvitopi}
\affiliation{%
    \institution{Lawrence Berkeley Nat. Laboratory}
    \city{Berkeley}
    \state{CA}
    \postcode{94720}
    \country{USA}}
\email{roselvitopi@lbl.gov}
\author{Israt Nisa}
\affiliation{%
    \institution{Amazon Web Services}
    \city{Palo Alto}
    \state{CA}
    \country{USA}}
\email{nisisrat@amazon.com}
\author{Zoran Budimli\'{c}}
\affiliation{%
    \institution{Rice University}
    \city{Houston}
    \state{TX}
    \country{USA}}
\email{zoran@rice.edu}
\author{Ayd{\i}n Bulu\c{c}}
\affiliation{%
    \institution{Lawrence Berkeley Nat. Laboratory}
    \city{Berkeley}
    \state{CA}
    \country{USA}}
\email{abuluc@lbl.gov}

\date{\today}

\usepackage{graphicx}
\usepackage{multirow}
\usepackage{multicol}
\usepackage{xcolor}
\usepackage[T1]{fontenc}
\usepackage{caption}
\usepackage{subcaption}
\usepackage{enumitem}
\usepackage{titlesec}

\usepackage{amsmath}
\usepackage{amsfonts}
\usepackage{color}
\usepackage{wrapfig}
\usepackage{algorithm}
\usepackage[noend]{algpseudocode}
\makeatletter
\renewcommand{\ALG@beginalgorithmic}{\footnotesize}
\makeatother

\titlespacing*{\section}
{0pt}{5pt plus 0pt}{3pt plus 0pt}
\titlespacing*{\subsection}
{0pt}{4pt plus 0pt}{2pt plus 0pt}

\newcommand{\InParallel}{\textbf{in parallel} }
\newcommand{\Break}{\textbf{break} }

\algdef{SE}% flags used internally to indicate we're defining a new block statement
[CLASS]% new block type, not to be confused with loops or if-statements
{Class}% "\Struct{name}" will indicate the start of the struct declaration
{EndClass}% "\EndStruct" ends the block indent
[1]% There is one argument, which is the name of the data structure
{\textbf{class} \textsc{#1}}% typesetting of the start of a struct
{\textbf{end class}}% typesetting the end of the struct

\newcommand{\ncols}{\mathit{ncols}}
\newcommand{\dnnz}{\mathit{nnz}}
\newcommand{\flops}{\mathit{flops}}

\newcommand{\mA}{\mathbf{A}}
\newcommand{\mB}{\mathbf{B}}
\newcommand{\mC}{\mathbf{C}}
\newcommand{\mM}{\mathbf{M}}

\newcommand{\mL}{\mathbf{L}}

\newcommand{\vu}{\mathbf{u}}
\newcommand{\vvv}{\mathbf{v}}
\newcommand{\vm}{\mathbf{m}}

\newcommand{\erdosrenyi}{Erd\H os-R\'{e}nyi}
\newcommand{\SSGB}{SS:GB}

\algrenewcommand\algorithmicrequire{\textbf{Input:}}
\algrenewcommand\algorithmicensure{\textbf{Output:}}

\newcommand{\sset}{\texttt{SET}}
\newcommand{\sall}{\texttt{ALLOWED}}
\newcommand{\snall}{\texttt{NOTALLOWED}} 

\algnewcommand\algorithmicforeach{\textbf{for each}}
\algdef{S}[FOR]{ForEach}[1]{\algorithmicforeach\ #1\ \algorithmicdo}

\setcopyright{none} 

\begin{document}
% \pagecolor[HTML]{333333} %% dark gray background
% \color[HTML]{CCCCCC} %% light gray foreground text

\begin{abstract}
Computing the product of two sparse matrices (SpGEMM) is a fundamental operation in various combinatorial and graph algorithms as well as various bioinformatics and data analytics applications for computing inner-product similarities. 
For an important class of algorithms, only a subset of the output entries are needed, and  the resulting operation is known as Masked SpGEMM since a subset of the output entries is considered to be ``masked out''. 

Existing algorithms for Masked SpGEMM usually do not consider mask as part of multiplication and either first compute a regular SpGEMM followed by masking, or perform a sparse inner product only for output elements that are not masked out.
In this work, we investigate various novel algorithms and data structures for this rather challenging and important computation, and provide guidelines on how to design a fast Masked-SpGEMM for shared-memory architectures.
Our evaluations show that factors such as matrix and mask density, mask structure and cache behavior play a vital role in attaining high performance for Masked SpGEMM.
We evaluate our algorithms on a large number of real-world and synthetic matrices using several real-world benchmarks and show that our algorithms in most cases significantly outperform the state of the art for Masked SpGEMM implementations.

		% In this paper, we give new shared-memory algorithms for Masked-SpGEMM that are based on the high-level logic of the row-by-row accumulator-based algorithms, but initialize and utilize the accumulator data structures in a way that exploits the mask structure to avoid redundant data access and/or writes.   
		
	\end{abstract}
	
\maketitle
\pagestyle{plain}

\section{Introduction}
Masked sparse-sparse matrix multiplication (\emph{Masked SpGEMM}) is the problem of computing the product of two sparse matrices only for the set of entries given by the nonzero structure of the mask. The mask can be thought as a sparse matrix whose pattern determines which elements should exist in the output matrix. While the first use of this primitive was in the context of triangle counting~\cite{azad2015parallel}, its applications include any multi-source graph traversal where the mask serves as a filter to avoid rediscovery of previously discovered vertices. A canonical example is the multi-source betweenness centrality as implemented in GraphBLAS C API~\cite{bulucc2017design}. Recently, Etter et al.~\cite{etter2021accelerating} showed how to accelerate tree-based inference methods using masked SpGEMM.
	
\begin{figure}
	\centering
	\includegraphics[width=0.75\columnwidth]{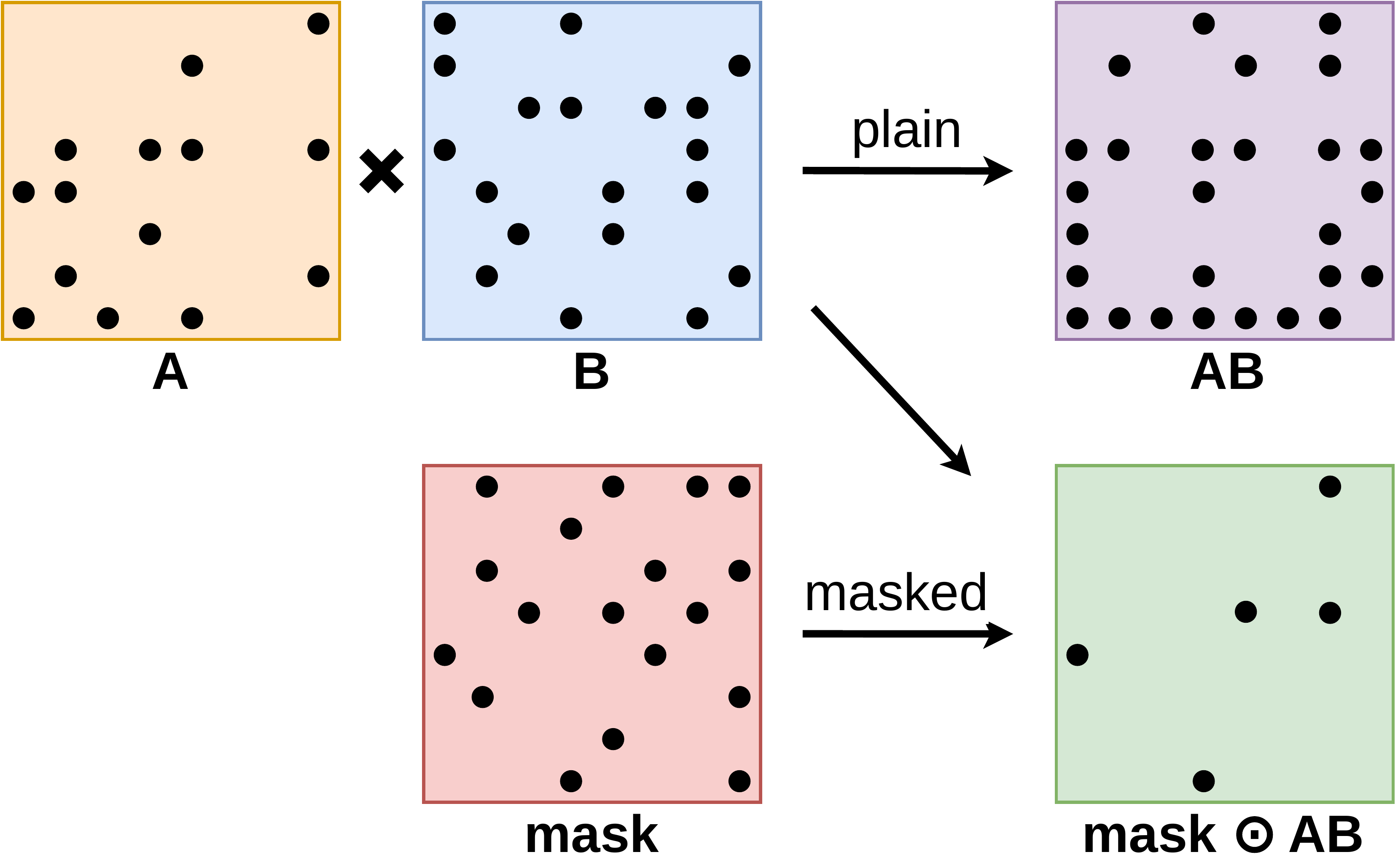}
	\caption{Plain vs. Masked SpGEMM. The masked output entries need not be computed. Also, mask may contain entries for which the multiplication does not produce an output.}
%	\vspace{-2ex}
	\label{fig:plain-vs-masked}
\end{figure}

%% PAR: vs. spgemm
The existence of a mask in the multiplication introduces new optimization opportunities as well as challenges. 
A simple way to perform Masked SpGEMM is to compute the multiplication as if the mask does not exist and then apply the mask to the output matrix, which
 causes unnecessary computation if the overlap between the output matrix and the mask is low (see Figure~\ref{fig:plain-vs-masked}).
The mask needs to be considered as part of the multiplication to attain good performance, which is the focus of this work.
%
%This work is an effort in that direction and examines the best practices developed over the years for SpGEMM when a mask is involved in the multiplication. 

%% PAR why need research on masked spgemm
Most parallel SpGEMM methods rely on Gustavson's algorithm~\cite{gustavson1978two}, in which a row or a column of the output matrix is computed by accumulating the partial results produced by scaling rows or columns of one of the input matrices.
The important design aspects of this algorithm, such as parallelization granularity, data structures (i.e., accumulators) used in the merging of partial results or whether to include a symbolic multiplication phase to determine the pattern of the output matrix, need to be reconsidered when a mask is part of the equation, even calling into question the viability of this algorithm for certain cases.
Consider the computation of a row of the output matrix in which a considerable amount of flops is spent to get the result.
If the mask does not require most of the entries in that row of the output matrix, one can avoid unnecessary computations by computing the unmasked entries with inner products instead of accumulating the scaled rows.
Moreover, many graph algorithms rely on operations involving the complement of the mask, which is a way to express avoiding already visited nodes.
This adds another design and optimization dimension to the Masked SpGEMM.
Hence, not only the specific details of established SpGEMM algorithms must be reexamined for Masked SpGEMM, but also the viability of other less frequently-utilized algorithms and new issues arising because of masking.

Our code implementing our algorithms and data structures is available at \href{https://github.com/PASSIONLab/MaskedSpGEMM}{https://github.com/PASSIONLab/MaskedSpGEMM}.

Our contributions in this paper are:
\begin{itemize}[topsep=0pt,itemsep=-1ex,partopsep=1ex,parsep=1ex,leftmargin=*]
    \item We describe push- and pull-based algorithms for Masked SpGEMM, and analyze/compare their memory behaviors. %The push-based Masked SpGEMM relies on Gustavson's algorithm, whereas the pull-based Masked SpGEMM relies on sparse dot products.
    \item We design four different data structures to be used as accumulators in Masked SpGEMM: (i) Hash, (ii) Masked Sparse Accumulator, (iii) Masked Compressed Accumulator, and (iv) Heap. The Masked Compressed Accumulator is a novel accumulator we specifically designed for Masked SpGEMM, while the remaining three are enhancements to the accumulators utilized in plain SpGEMM.
    \item We discuss how to adapt these accumulators for Masked SpGEMM where the mask is complemented.
    \item In SpGEMM, a symbolic phase is often performed to compute the pattern of the output matrix prior to the numeric phase. We review the tradeoffs of including a symbolic phase when mask is part of the SpGEMM.
    \item We conduct extensive experiments on both synthetic and real-world matrices using graph processing applications to reveal the best design choices for a fast Masked SpGEMM.
\end{itemize}

%% PAR contributions
%% PAR rest
% \Oguz{UPDATE WHEN SECTIONS CHANGE.}
% The rest of this paper is organized as follows.
% %
% In Section~\ref{sec:bg} we give the necessary background and notation we use and in Section~\ref{sec:rw} we review the related work.
% %
% The pull- and push-based algorithms for Masked SpGEMM are described in Section~\ref{sec:pull_algs} and Section~\ref{sec:push_algs}, respectively.
% %
% The generic design of an accumulator for Masked SpGEMM and four different data structures proposed for that purpose are described in Section~\ref{sec:algs}.
% %
% Section~\ref{sec:phase} discusses symbolic and numeric phases for Masked SpGEMM.
% %
% Section~\ref{sec:analysis} analyzes and compares Masked SpGEMM algorithms discussed in the paper. \Zoran{We don't have an analysis section yet}
% %
% Finally, Section~\ref{sec:exp} presents our experimental evaluation and Section~\ref{sec:conc} concludes.

% When the input matrices to be multiplied are dense, the masked multiplication is known as the sampled dense-dense matrix multiplication (SDDMM)~\cite{canny2013big}, and has applications in machine learning such as matrix factorization~\cite{nisa2018sampled} and graph neural networks~\cite{hu2020featgraph}. The sparsity pattern of the output matrix in known beforehand in SDDMM unlike SpGEMM.

\section{Background and Notation}
\label{sec:bg}
%% PAR graph <-> matrix, semiring and GraphBLAS
SpGEMM forms the computational backbone of many applications in linear algebra~\cite{Bell2012, Briggs2000, Griewank2003} and graph processing~\cite{Buluc2012, Gilbert2008, Penn2006, Dongen2008, azad2015parallel, Wolf2017, Selvitopi2020}.
This paper targets the performance of masked SpGEMM for graph processing as it is where the masked variant of SpGEMM is mainly utilized.
Many graph algorithms can be expressed in terms of computations on sparse matrices due to the duality between graph and matrices.
%
%However, departing from the sparse computations in linear algebra, the operations on graphs require capabilities unique to graph algorithms.
%
%For example, a graph algorithm may want to associate custom data with nodes and edges of the graph or it may want to access neighbors of certain nodes and perform non-arithmetic operations on them.
%
In this direction, GraphBLAS~\cite{bulucc2017design} is an effort to standardize the graph algorithm primitives in the language of linear algebra with an extended set of algebraic objects including semirings, upon which common sparse matrix computations such as SpMV or SpGEMM can be generalized.

%% PAR notation
We denote the Masked SpGEMM with $\mC = \mM \odot (\mA \mB)$ on a semiring, 
%$S=\langle D_A, D_B, D_C, \oplus, \otimes, \mathrm{\texttt{id}} \rangle$, 
where $\mM \in \mathbb{S}^{m\times n}$ is the mask, $\mA \in \mathbb{S}^{m\times k}$ and $\mB \in \mathbb{S}^{k\times n}$ are the input sparse matrices, and $\mC \in \mathbb{S}^{m\times n}$ is the output sparse matrix.
%
%In the semiring, $D_A$, $D_B$, and $D_C$ respectively denotes the data types of the matrices $A$, $B$, and $C$, $\oplus$ is the binary additive operation, $\otimes$ is the binary multiplicative operation, and \texttt{id} is the identity element of the semiring.
%
%We prefer the GraphBLAS semirings~\cite{bulucc2017design} instead of conforming to the requirements of a mathematical semiring in order to allow more flexibility in expressing graph operations.
%
%When the data types used for the SpGEMM are irrelevant to our discussions, we may use $S=\langle \oplus, \otimes, \mathrm{\texttt{id}} \rangle$.
%
Although the graph algorithms evaluated in this work utilize various semirings, we use the arithmetic semiring in our algorithms to keep the discussions simple.
$\mA_{ij}$ denotes a nonzero element of $\mA$, and $\mA_{i*}$ and $\mA_{*j}$ respectively denotes $i$th row and $j$th column of $\mA$.
Note that we only utilize the pattern of the mask in Masked SpGEMM, hence the values in the mask are not evaluated and the type of the mask elements does not matter.
We denote a vector with lowercase letters, i.e., $\vvv$, and an element of a vector is denoted with $\vvv_i$.
We use the function $\dnnz(\cdot)$ to denote the number of nonzero elements in a matrix/vector, $\ncols(\cdot)$ to denote the number of columns in a matrix/vector, and $\flops(\cdot)$ to denote the number of floating point operations in a sparse matrix operation.
%, and $\size(\cdot)$ to denote the size of a vector or an array.

\begin{algorithm}[t]
	\caption{Row-parallel Gustavson SpGEMM~\cite{gustavson1978two}}
	\label{algo:code_spgemm}
	\begin{algorithmic}[1]
	\Require Sparse matrices $\mA$, $\mB$%, Semiring $S=\langle \oplus, \otimes, \mathrm{\texttt{id}} \rangle$
	\Ensure Sparse matrix $\mC$
		\State $\mC \leftarrow \emptyset$
		\ForEach{row $\mA_{i*}$ in matrix $\mA$ \InParallel}
		\ForEach{nonzero $\mA_{ik}$ in row $\mA_{i*}$}
		\ForEach{nonzero $\mB_{kj}$ in row $\mB_{k*}$}
		\State $value \leftarrow \mA_{ik} \cdot \mB_{kj}$
		\If {$\mC_{ij} \not\in \mC_{i*}$}
		\State insert $(\mC_{ij} \leftarrow value)$ to $\mC_{i*}$
        \Else
		\State $\mC_{ij} \leftarrow \mC_{ij} + value$
		\EndIf
		\EndFor
		\EndFor
		\EndFor
	\end{algorithmic}
\end{algorithm}

%% PAR cache 
In our analyses, we assume that $\dnnz(\mA), \dnnz(\mB), \dnnz(\mM) \gg Z$, where $Z$ is the size of the last-level cache.
For simplicity, we assume that a cache line can hold $L$ words, and that integers used in indexing and values used to store data are all the same size of a single word. 
%
%Since each memory transfer happens in blocks of $L$ words, fully random %accesses increase the memory bandwidth by a factor of $L$ compared to the %amount of useful data requested. 

\subsection{Storage Formats}
%% PAR formats used
Most popular formats for storing sparse matrices are Compressed Sparse Row (CSR), Compressed Sparse Column (CSC), Coordinate, and Doubly-Compressed Sparse Row/Column (DCSR/DCSC).
In this work, we use the CSR format in most cases, with CSC only being used in a single case to improve performance of the inner product. We use CSR format for storing the mask.
The CSR (CSC) format uses three arrays to store a sparse matrix: an array containing row (column) index pointers, an array containing column (row) indices of nonzeros, and an array containing values of nonzeros.

%% PAR mask sparse vs. dense
%We note that in Masked SpGEMM the mask has no requirement to be sparse.
%
%Indeed, some graph algorithms may utilize a dense mask as part of their execution.
%
%(even when it is dense). 
%as it is relatively straightforward to handle a mask when it is stored as dense.
%
%We exclude other storage formats as well as the possibility of storing the mask as a dense matrix from our discussions since we believe the issues investigated in this work are more general and have a higher importance.

\subsection{Design Issues and Challenges}
%% PAR challenges
There are four challenges to designing an efficient parallel SpGEMM running on multi-core systems: (i) irregular and random memory accesses when retrieving rows or columns of a sparse matrix, (ii) designing an accumulator to merge the partial results (if any), (iii) determining the pattern of the output matrix, and (iv) load imbalance.
%
%These factors are not independent of each other and a choice made in one of them may affect others.
%
%
SpGEMM algorihms usually access the rows or columns of the sparse matrices randomly and this causes SpGEMM operation to be memory-bound rather than compute-bound, and often results in poor cache behavior. The addition of a mask to the SpGEMM only exacerbates these challenges.

%% PAR accumulator
A large portion of SpGEMM algorithms necessitate a data structure to accumulate the partial results to get the output matrix entries.
For example in algorithms that rely on Gustavson's algorithm~\cite{gustavson1978two}, they may be used to accumulate the products in scaled rows/columns that correspond to the same output entry (Algorithm~\ref{algo:code_spgemm}).
%
%Note that it may not be always necessary to use an accumulator, such as in an inner-product-based algorithm or in an expansion-sort-compress-based method~\cite{Bell2012, Dalton2015}.
%
%These, however, are usually noncompetitive against the algorithms that rely on accumulators.
%
Among common accumulators are sparse arrays~\cite{gilbert1992sparse, Patwary2015} (SPA), heaps~\cite{Azad2016, Liu2014}, and hash tables~\cite{Deveci2017, Nagasaka2017}.
We investigate how to enhance these accumulators for Masked SpGEMM and propose a novel accumulator specifically designed for Masked SpGEMM.

%% PAR 1-phase vs. 2-phase
Another issue is the unknown size and structure of the output matrix, which renders memory management difficult for the output matrix.
Thus, SpGEMM algorithms sometimes perform two passes to complete the multiplication: the first pass to figure out the size and structure of the output matrix, referred to as the \emph{symbolic} phase, and the second pass to compute its entries, referred to as the \emph{numeric} phase.
These are known as the \emph{two-phase} algorithms, as opposed to the \emph{one-phase} algorithms which allocate a large-enough memory at the beginning and perform the multiplication in a single numeric phase.
%
%Note that it is impossible to know the memory required by the output matrix beforehand, so the ``large-enough memory'' is anyone's guess.
%
Because the pattern of the mask partially determines the structure of the output matrix (note that the mask may contain unmasked entries for which there is no output entry), so one can possibly utilize the mask as a good starting point to approximate the size and the structure of the output. This can potentially render two-phase algorithms obsolete, a point we investigate in this work.
%
%This is one of the main investigation points of this work as we include both a one-phase and a two-phase variant of each of the Masked SpGEMM algorithms proposed in this work.
% \Aydin{Aydin writes}
% Consider the operation $\mC = \mM \odot (\mA \mB)$ where $\mM \in \mathbb{S}^{m\times n}$ is the mask, $\mA \in \mathbb{S}^{m\times k}$ and $\mB \in \mathbb{S}^{k\times n}$ are the input sparse matrices. 

\section{Related Work}
\label{sec:rw}
SuiteSparse:GraphBLAS (\SSGB{}, which is the fastest compliant GraphBLAS implementation on multicore processors, initially used a dot-product algorithm for most masked SpGEMM calls~\cite{aznaveh2020parallel}.Its most recent version~\cite{ssgrb21} also implements various hash-based and SPA-based codes that exploit mask. %where the mask is probed with a binary search before writing to the output. 
As a complete library, \SSGB{} supports multiple sparse matrix data structures, concepts such as ``pending tuples'' and ``zombies'' to take advantage of lazy evaluation in non-blocking mode of GraphBLAS, and graph specific optimization such as iso-valued matrices where all nonzero entries in the matrix have the same value. Therefore, it is not reasonable to perform an apples-to-apples comparison with our work, which focuses on algorithmic differences among various methods for performing masked SpGEMM. We are instead going to highlight the differences between our masked SpGEMM algorithms and those implemented in the latest version of \SSGB{}.

The \texttt{GrB\_mxm} function of SuiteSparse:GraphBLAS, which covers the case of Masked SpGEMM, works with 4 different matrix formats: sparse, hypersparse, bitmap, and dense. The sparse case uses either the CSR or CSC formats. The hypersparse case uses either DCSR or DCSC~\cite{buluc2008representation}. Our work is focused on the CSR format to isolate the algorithmic trade-offs involved in Masked SpGEMM. Our algorithms do not parallelize the formation of individual rows as we observed that there is plenty of coarse-grained parallelism across rows to avoid any load imbalance on the multi-core processors available today. The only other Masked SpGEMM implementation we are aware of is the GPU implementation from the GraphBLAST library~\cite{yang2019graphblast}, which is based on dot products.

% \Aydin{The following is Tim's version of our differences: Milaković et al., [23] in work done independently of the work reported in this paper, have developed four novel parallel algorithms for the masked matrix-matrix multiply, C⟨M⟩ = AB and C⟨¬M⟩ = AB where all input matrices are sparse (not hypersparse, bitmap, or full). These include a (1) hash-based method similar to the masked coarse Hash method described in Section 4.2, and (2) a masked sparse accumulator (MSA) method, similar to the coarse Gustavson method in Section 4.2. They describe two algorithms mostly unrelated to the methods described in this paper: (3) a heap-based method, and (4) a mask compressed accumulator method (MCA). The first three methods support both a complemented and non-complemented mask, while the MCA method only supports a non-complemented mask. The traversal of the mask in the MCA method is analogous to traversing M(:,j) and using a binary search of A(:,k) discussed in Section 4.2.
	
% The fine-grain methods presented in this paper are unique to this paper, and not considered by Milaković et al., as are the combinations of sparse/hypersparse/bitmap/full considered here, and the exploitation of the iso properties of the four matrices C, M, A, and B. In SuiteSparse:GraphBLAS, a single matrix multiply can use any mix of four kinds of tasks (coarse/fine Hash/Gustavson), which is also unique to this paper.}

\section{Classification of Algorithm Families} 
Our work on masking out certain entries of the output in sparse matrix computation has its primary motivations in the area of graph processing. In particular, the concept of masking has been first applied to sparse-matrix-vector multiplication to implement the direction-optimized graph traversal~\cite{yang2018implementing}. The direction-optimization~\cite{beamer2012direction} is also known as push-pull~\cite{besta2017push} in the graph processing community. The standard way of processing a graph involves a frontier of active vertices that ``pushes'' information to their out-neighbors. The pull operation happens when the non-active (or previously unvisited, depending on the application) vertices ``pulling'' information from their in-neighbors. 

This analogy with graph processing allows us to provide a classification of Masked SpGEMM algorithms into two main classes. The pull-based algorithms are those whose computation is mainly driven by the mask. For each potential entry $\mC_{ij}$ in the output that is not masked out (i.e., $\mM_{ij} \neq 0$), we ``pull'' information by inspecting the input sparse matrices to see if they can generate that output entry. The push-based algorithms are instead driven by the sparsity of the input matrices first, and they often utilize the mask as a filter before generating the output. 

%Below we provide a high-level comparison of these families of algorithms in terms of their parallelism and cache utilization. 

\subsection{Pull-based Algorithms}
\label{sec:pull_algs}
Consider the na\"ive algorithm where for each $\mM_{i,j} \neq 0$, we perform the sparse dot product $\mA_{i*} \mB_{*j}$. 
Since such sparse dot products are independent of each other, this algorithm has at least $O(\dnnz(\mM))$-way parallelism, excluding any parallelism that can be extracted within the sparse dot product itself. 
%
%We call this algorithm the \emph{masked-na\"ive-pull} method. 
%
This method is most efficiently implemented when $\mA$ is stored in a row-major sparse storage such as CSR, and $\mB$ is stored in a column-major sparse storage such as CSC (or vice versa), which is what we assume is the case in this study.

The described method, however, has one drawback: its poor temporal locality. 
Assume that we traverse the nonzeros of $\mM$ in row-major order. 
Since rows of $\mA$ are accessed consecutively, there is a significant reuse within rows. 
However, columns of $\mB$ will be accessed in a scattered manner, with very little reuse. 
Given $\dnnz(\mB) \gg Z$ where $Z$ is the fast memory (cache) capacity, we can assume that each column access will fetch the whole column back from the main (slow) memory. 
For simplicity, we assume that each column of $\mB$ is accessed the same number of times or each column of $\mB$ has the same number of nonzeros $\dnnz(\mB)/n$. 
Either way, the amount of memory traffic of this method is:
$\dnnz(\mA) +  \dnnz(\mM) \Big( 1 + \frac{\dnnz(\mB)}{n} \Big).$
\subsection{Push-Based algorithms}
\label{sec:push_algs}

There are many push-based SpGEMM algorithms. 
In this section, we focus on the most well-known row-by-row version due to Gustavson~\cite{gustavson1978two}, also shown in Algorithm~\ref{algo:code_spgemm}.
In this algorithm, the $i$th row of the output is computed as a linear combination of $k$th rows of $\mB$ where $\mA_{ik} \neq 0$. %
This algorithm naturally parallelizes over rows as there are no dependencies.
However, using a sparse accumulator (SPA) increases the cache load as one dense vector per row is used for accumulation. 
To overcome this, researchers have used data structures ranging from priority queues to hash tables to merge sparse rows of $\mB$.

% \begin{figure}
% 	\centering
% 	\includegraphics[width=\columnwidth]{GustavsonSPAPicture.pdf}
% 	\caption{Column-wise sparse-matrix multiplication using SPA (sparse accumulator). All matrix columns and vectors are stored compressed except the SPA itself.}
% 	\label{fig:SPAGustavson}
% \end{figure}

The formation of $\mC_{i*}$ exhibits five memory access patterns:
\begin{enumerate}[topsep=0pt,itemsep=-1ex,partopsep=1ex,parsep=1ex,leftmargin=*]
\item Unit-stride read access to nonzeros within a row of $\mA$. 
	\item Random-like read access to row pointers of $\mB$.
	\item Stanza-like read access to nonzeros of $\mB$: small blocks (stanzas) of consecutive elements are fetched from effectively random locations in memory. 
	\item Random-like read and write access to the sparse accumulator (the scatter/accumulate step) for updating values.
	\item Unit-stride write access for outputting $\mC_{i*}$. 
\end{enumerate}

The choice of the data structure only changes the $4$th type of access where we can improve cache utilization with more compact data structures. 
The first three memory access patterns are canonical for row-by-row algorithms, i.e., they persist even if we replace the SPA with a priority queue or a hash table.
As long as we use the push paradigm, those three memory accesses are also not affected by the use of a mask.

To analyze memory traffic costs, we make two reasonable assumptions: (1) the cache line length $L$ is smaller than the matrix dimension, i.e. $n > L$, and (2) the bandwidth of the first matrix is larger than the cache size, i.e. $\beta(\mA) > Z$. The matrix bandwidth $\beta(\mA)$ is the smallest non-negative integer $k$ such that $\mA_{ij} \neq 0$ for $\lvert i - j \rvert > k$. Corollaries of Assumption 2 when performing $\mA \mB$ using a row-by-row algorithm are (a) row pointers of $\mB$ are not cached, and (b) accesses to column ids and values of distinct rows of $\mB$ are not cached. 

The memory traffic incurred due to the first pattern is trivially $O(\dnnz(\mA))$. The memory cost of the second pattern is $O(\dnnz(\mA) \cdot L)$ due to assumption 2. The third pattern incurs $O(\flops (\mA \mB))$ memory traffic, again due to the assumption 2. We do not analyze the last two steps in this section because they depend on whether we use the mask or not, and what particular data structure we use to store the mask.

\subsection{High-level Comparison}
Let us use fixed input sparsity $d=\dnnz(B)/n = \dnnz(A)/n$ for the sake of comparison. When both the input matrices get denser, the push-based row-by-row algorithms gets expensive quadraticly with $d$, because in that case $O(\flops (\mA \mB)) = d^2 n$. However, pull-based dot-product algorithm gets expensive only linearly with $d$.

On the other hand, when the mask gets asymptotically sparser than the input, say for $d_m=  \dnnz(M)/n \lll d $, the pull-based algorithms tend to outperform push-based algorithms, regardless of the choice of mask data structure.
\section{Our Algorithms}
\label{sec:algs}

\begin{figure*}
	\centering
	\includegraphics[width=0.85 \textwidth]{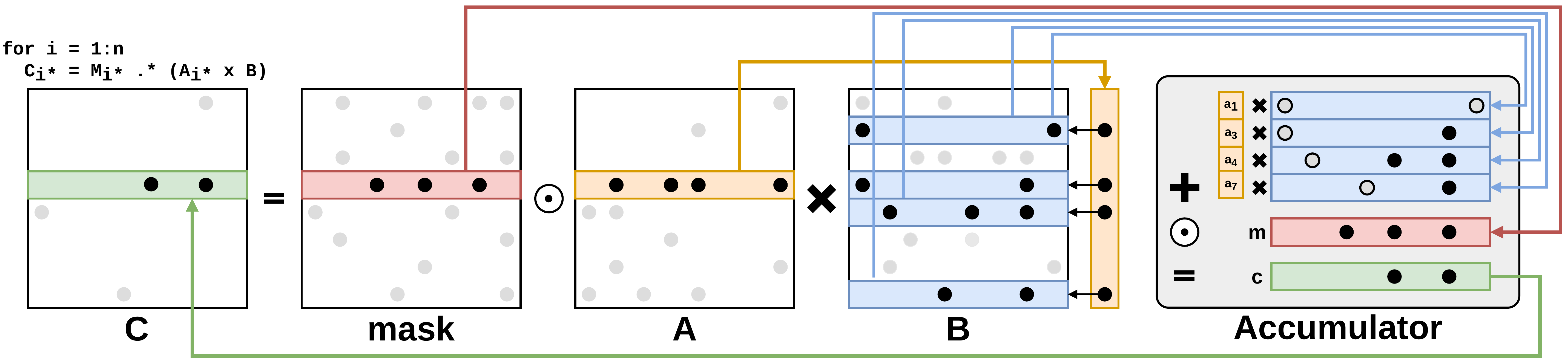}
	\caption{Row-wise Masked SpGEMM using an accumulator to compute output row  $\mC_{i*}$. The rows corresponding to the column indices of entries in row $\mA_{i*}$ are merged and filtered through the respective mask entries to compute $\mC_{i*}$. This merging and filtering process can be performed in a number of ways.}
	%\vspace{-.1in}
	\label{fig:masked-spgemm}
\end{figure*}

%\Oguz{I think we should also describe the inner product algorithm as it is evaluated in the experiments.}

We describe four novel algorithms for Masked SpGEMM: Hash, Masked Sparse Accumulator (MSA), Mask Compressed Accumulator (MCA), and Heap.
Three of these algorithms --Hash, MSA, and Heap-- are novel improvements to the SpGEMM algorithms described in \cite{deveci2018multithreaded, nagasaka2019performance, buluc2008representation}, whereas MCA, to the best of our knowledge, is a completely new algorithm specifically developed for Masked SpGEMM.
All algorithms belong to the category of row-by-row push-based algorithms (Section~\ref{sec:push_algs}).
The computational flow of row-by-row Masked SpGEMM algorithms is illustrated Figure~\ref{fig:masked-spgemm}.
Each row $\mC_{i*}$ is calculated as element-wise multiplication of $\mM_{i*}$ and linear combination of each row $\mB_{k*}$ for which $\mA_{ik} \neq 0$, i.e., $\mC_{i*}=\mM_{i*} \odot \sum_{\mA_{ik} \neq 0}{\mA_{ik}  \mB_{k*}}$.
Notice that calculation of each row can be seen as a row vector-matrix multiplication (SpGEVM) followed by mask operation $\vvv^\intercal = \vm^\intercal \odot \left( \vu^\intercal \mB \right)$. 
%\Oguz{need to make clear whether we use row or column vectors, and stick to it in this chapter}
%
In order to simplify the algorithm explanations, without loss of generality, in this section we describe different algorithms for computing Masked SpGEVM.
Extrapolating Masked SpGEVM algorithms to devise Masked SpGEMM algorithms is straightforward.

\subsection{Accumulator}
\label{sec:accum}
%In this section we explain accumulator role in Masked SpGEVM, and the methods that the accumulator interface contains.
%Before describing the masked SpGEVM algorithm, we define the accumulator interface and its role.
%Accumulator is a data structure used for generating the resulting vector $\vvv$ by accumulating products $\vu_{k}\mB_{kj}$ that have same column id in the output vector.

%Before describing the masked SpGEMM algorithm, we define the accumulator interface and its role.
%Accumulator is a data structure used for generating a single row $r$ in the resulting matrix $\mC$.
%The accumulator accumulates products $\mA_{rk}\mB_{kj}$ with same column index and discards all values $\mC_{rj}$ for which $\mM_{rj} = 0$. To reduce the number of unnecessary operations, the accumulator may not calculate product $\mA_{rk}\mB_{kj}$ that will be discarded. That is, the accumulator can calculate only products $\mA_{rk}\mB_{kj}$ for which $\mM_{ij} \neq 0$.

A key component in all the Masked SpGEVM algorithms is the {\em accumulator}, which is basically a data structure to merge scaled rows and can be considered as a set union operation.
The design and the implementation of the accumulator has a significant impact on memory hierarchy behaviour and therefore on the performance 
%\Oguz{and memory hierarchy behaviour}
of Masked SpGEVM, and is the key differentiating feature between our proposed algorithms, so a more detailed description of the accumulator warrants its own subsection.
We describe the accumulator as a generic interface that can be implemented differently by using various data structures in order to generate an output vector $\vvv$ (i.e., a row of $\mC$). %(a single row $r$ in the resulting matrix $\mC$).

%% PAR accumulator details
The accumulator accumulates products $\vu_k\mB_{kj}$ with the same column index $j$ and discards all the values $\vvv_{j}$ for which $\vvv_{j} = 0$. 
In order to reduce the number of unnecessary operations, the accumulator can altogether skip calculating the products that will be discarded. 
That is, the accumulator may calculate only the products $\vu_{k}\mB_{kj}$ for which $\vm_{j} \neq 0$.
Consequently, a more complex data structure than the Sparse Accumulator of Gilbert et al.~\cite{gilbert1992sparse} is needed. 
In particular, an accumulator for Masked SpGEVM needs to be able to differentiate between three states: \sset, \sall, and \snall. 

Our accumulator interface contains three procedures: 
%\textproc{setAllowed(key)}, \textproc{insert(key, value)}, and \textproc{remove(key)}.
\begin{itemize}[topsep=0pt,itemsep=-1ex,partopsep=1ex,parsep=1ex,leftmargin=*]
\item \textbf{\textproc{setAllowed(key)}} marks the values that should not be discarded and have the potential to be in the output matrix.
\item \textbf{\textproc{insert(key, value)}} inserts a key-value pair into the accumulator.
Since the value that is being inserted may be discarded, the insert procedure allows the second argument (value) to be a lambda function that will only be evaluated if the value it computes will not be discarded.
\item \textbf{\textproc{remove(key)}} accumulates all previously inserted values with the specified key and returns the value of the corresponding key. 
If no values with the specified key were previously inserted, or if  \textproc{setAllowed} is never called for the specified key, the procedure \textproc{remove} returns \textbf{none}. 
After the function \textproc{remove} is called, all values with the specified key are removed from the accumulator.
%and values with the specified key are marked as discardable. NOT TRUE for MCA
\end{itemize}

We next describe how the four accumulators implement this interface and how they are used to perform SpGEVM.
\subsection{Masked Sparse Accumulator (MSA)}
\label{sec:acc_msa}
\begin{figure}[h]
	\centering
	\includegraphics[width=0.35\textwidth]{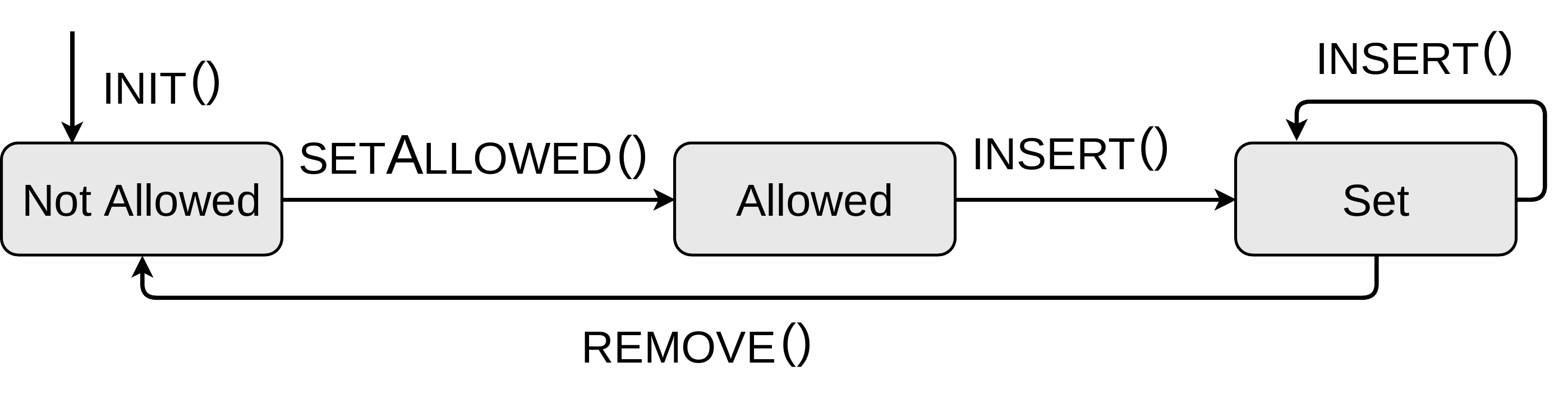}
	\caption{MSA states automaton}
   % \vspace{-.1in}
	\label{fig:msa-states}
\end{figure}

%In this section we describe masked sparse accumulator design (MSA) and masked SpGEVM algorithm that uses MSA.
%MSA implements the accumulator interface described in the previous section.
%
Internally, MSA uses two dense arrays, $values$ and $\mathit{states}$, each with $\ncols(\vvv)$ length.
$\mathit{values}$ stores the accumulated values, and $\mathit{states}$ stores information about the state of the entries in $\mathit{values}$, which may be one in of three states \snall{}, \sall, and \sset.
Initially, all of them are in the \snall{} state and the only valid transition from this state is to the \sall{} state, which is achieved by \textproc{setAllowed}.
%
%The only valid transition from \textrm{NotAllowed} is transition to state \textrm{Allowed}, and it is done via \textproc{setAllowed} procedure. %A state is changed from \textit{NotAllowed} to \textit{Allowed} via \textproc{setAllowed} procedure.
\textproc{insert} changes the state from \sall{} to \sset{}.
When key-value pair $(K, V)$ is inserted into the MSA, if key $K$ was previously marked as \sall{} and if no values with key $K$ were previously inserted, the respective is updated as \sset{} and $\mathit{values}_{\mathit{key}}$ is set to $V$.
Otherwise, if key $K$ was previously marked as \sset{} --meaning that some value with key $K$ was previously inserted-- the value $V$ is accumulated with the previous result.
\textproc{remove} returns the accumulated value with the specified key if the accumulated value was previously set and returns \textit{none} otherwise.
Figure \ref{fig:msa-states} shows MSA state automaton.
% Algorithm~\ref{algo:msa_accumulator} presents the pseudo-code for these operations.
%

% \Oguz{might be a good idea to include a simple figure with 3-state automata with states NotAllowed, Allowed, and Set, and perhaps remove Alg 2}

\begin{figure}[ht]
	%	\centering
	\begin{subfigure}[b]{0.155\textwidth}
		\centering
		\includegraphics[width=\textwidth]{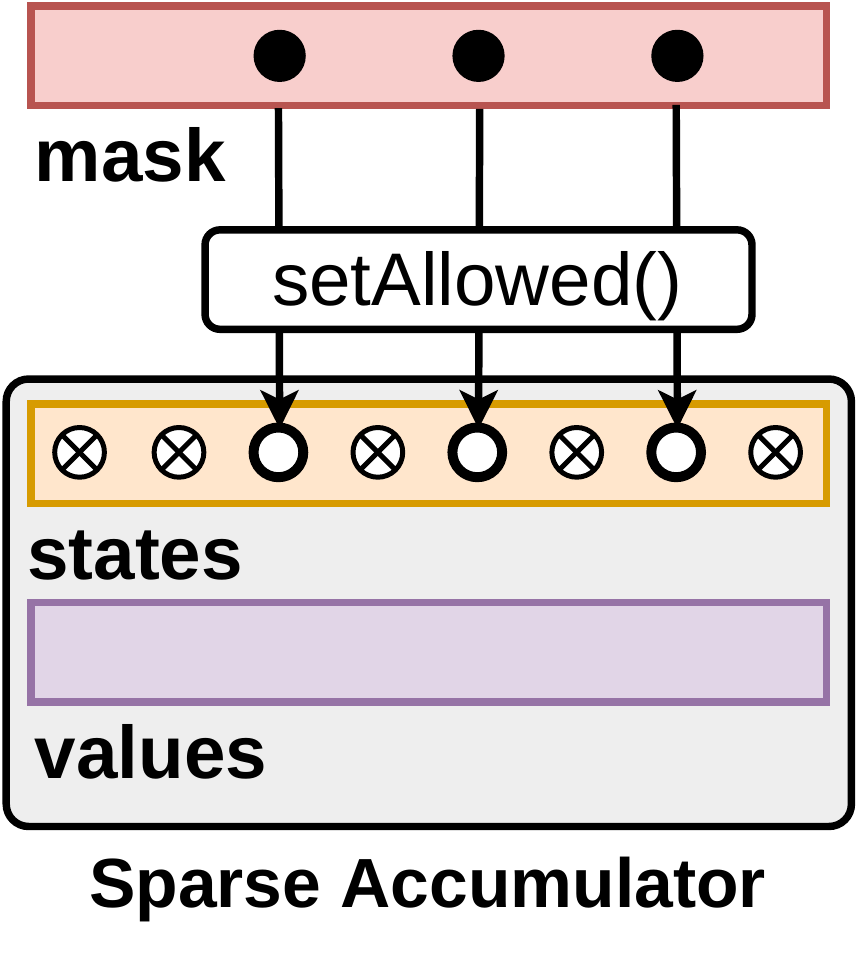}
		\caption{Set allowed keys}
		\label{fig:msa-alg-1}
	\end{subfigure}
	\begin{subfigure}[b]{0.155\textwidth}
		\centering
		\includegraphics[width=\textwidth]{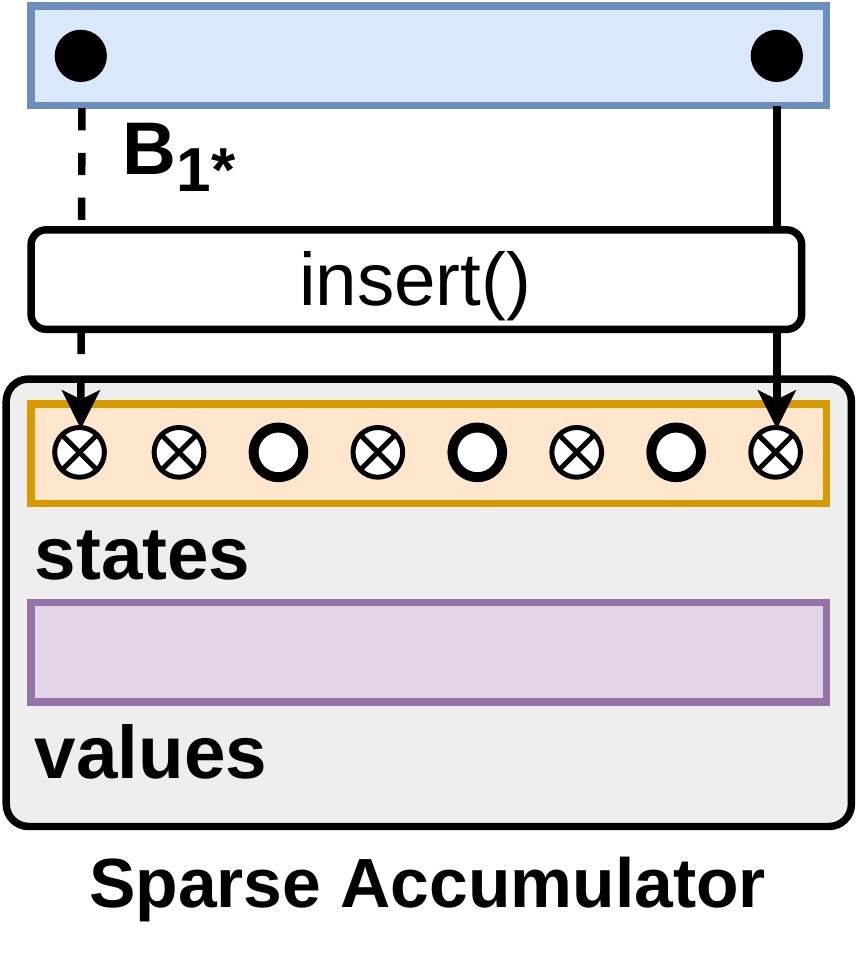}
		\caption{MSA += $\vu_1 \times \mB_{1*}$}
		\label{fig:msa-alg-2}
	\end{subfigure}
	\begin{subfigure}[b]{0.155\textwidth}
		\centering
		\includegraphics[width=\textwidth]{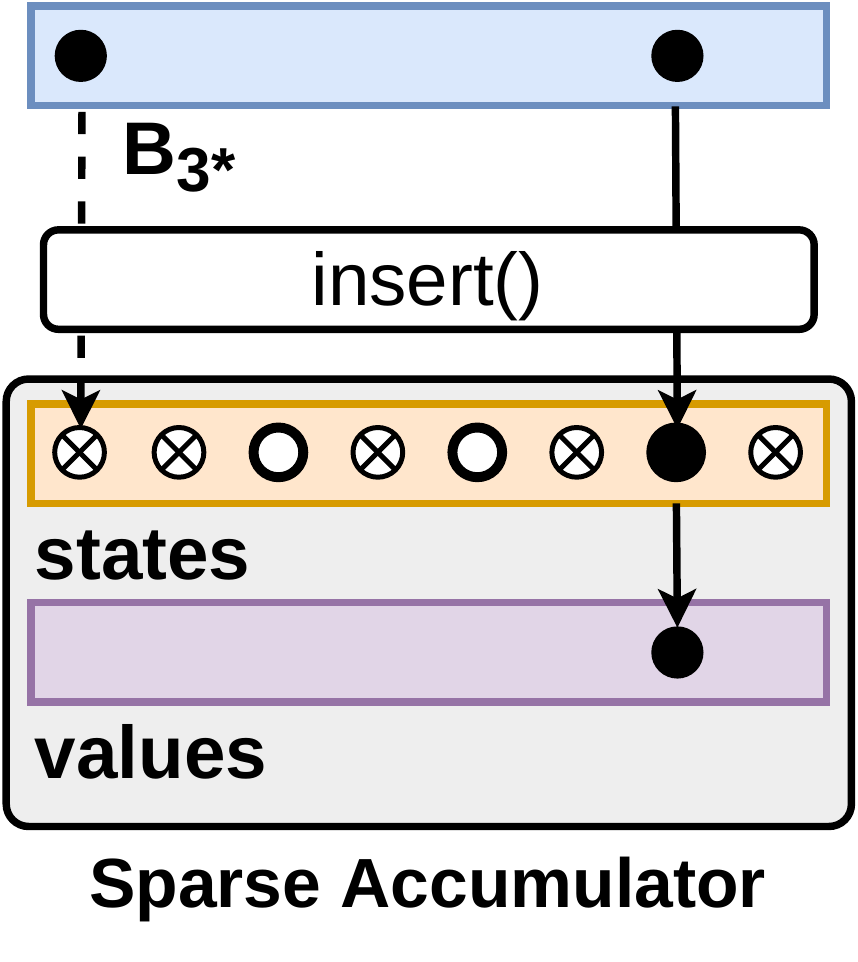}
		\caption{MSA += $\vu_3 \times \mB_{3*}$}
		\label{fig:msa-alg-3}
	\end{subfigure}
	\begin{subfigure}[b]{0.155\textwidth}
		\centering
		\includegraphics[width=\textwidth]{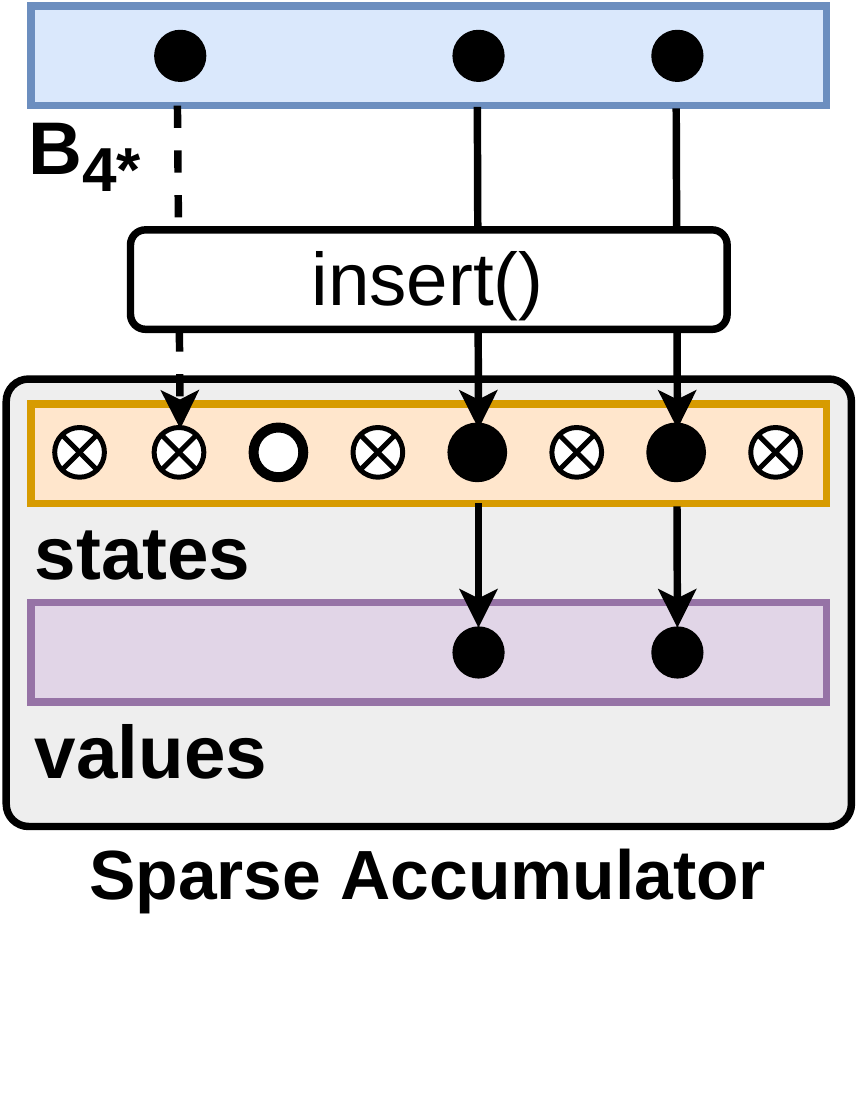}
		\caption{MSA += $\vu_4 \times \mB_{4*}$}
		\label{fig:msa-alg-4}
	\end{subfigure}
	\begin{subfigure}[b]{0.155\textwidth}
		\centering
		\includegraphics[width=\textwidth]{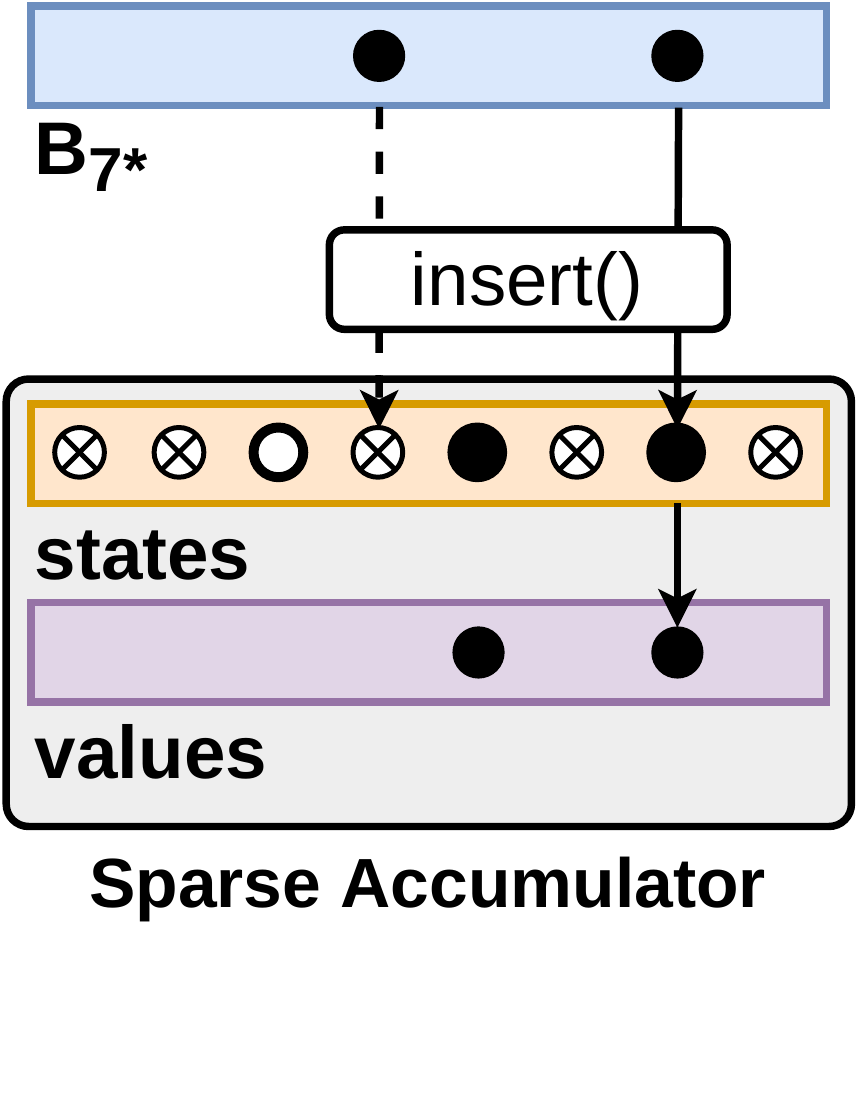}
		\caption{MSA+= $\vu_7 \times \mB_{7*}$}
		\label{fig:msa-alg-5}
	\end{subfigure}
	\begin{subfigure}[b]{0.155\textwidth}
		\centering
		\includegraphics[width=\textwidth]{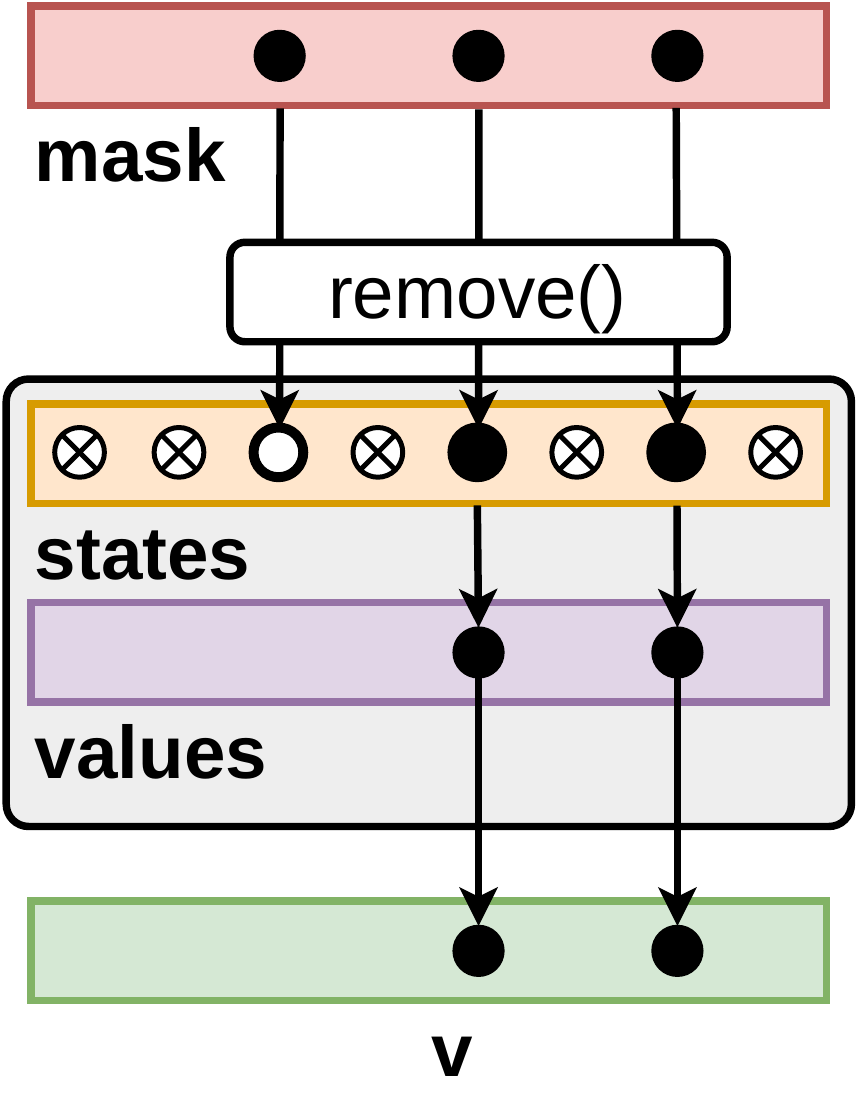}
		\caption{Move values to $\vvv$}
		\label{fig:msa-alg-6}
	\end{subfigure}
	\caption{Masked SpGEVM algorithm using Masked Sparse Accumulator (MSA).}
	%\vspace{-.1in}
	\label{fig:msa-alg}
\end{figure}

SpGEVM algorithm that uses MSA as accumulator is given in Algorithm~\ref{algo:msa_masked_spgevm} and has three main steps.
First, the algorithm initializes MSA and by using the mask elements it marks the values in the MSA that should not be discarded as \sall{}.
Second, the algorithm finds all products $\vu_{k}\mB_{kj}$ for which $\vu_{k} \neq 0$ and $\mB_{kj} \neq 0$, and inserts them into MSA.
Finally, the algorithm gathers all nonzero values from MSA, and resets the states in MSA.
To improve the performance of this last gather operation, the algorithm only iterates over the entries that are present in the mask.
Additionally, the values are gathered in the same order as they are ordered in the mask.
This has the benefit of being stable: if, for example, the values in the mask are sorted by column indices, the values in the output vector $\vvv$ will also be sorted by column indices.
Figure~\ref{fig:msa-alg} shows masked SpGEVM algorithm that generates output vector $\vvv$ using MSA, where the mask acts a filter to determine which output entries are valid or invalid.

MSA algorithm can also be used if the mask is complemented.
%
%The main difference between complemented and non-complemented mask algorithm, is that 
When the mask is complemented, the default state for the accumulator becomes \sall{}, and for each element in the mask we invoke \textproc{setNotAllowed} instead of \textproc{setAllowed}.
Additionally, if the mask is complemented we have an additional array to keep track of the elements that were inserted into the accumulator. 
This allows us to gather the accumulated values without iterating through the whole array.
Similar strategy was used by Gustavson ~\cite{gustavson1978two}.

MSA initialization takes $O(\ncols(\vu))$ operations, and computing the output vector $\vvv$ takes $O(\dnnz(\vm) + \flops(\vu \mB))$ operations. 
Therefore, the algorithm takes $O(\ncols(\vvv) + \dnnz(\vm) + \flops(\vu \mB))$ operations.

% \begin{algorithm}[t]
% 	\caption{Masked Sparse accumulator (MSA)}
% 	\label{algo:msa_accumulator}
% 	\begin{algorithmic}[1]
% 		\Class{SparseAccumulator}
		
% 		\Procedure{init}{maxKey}
% 		\State \{allocate dense arrays for states and values\}
% 		\State $states_* \leftarrow \snall$
% 		\EndProcedure
% 		%		\State
% 		\Procedure{setAllowed}{key}
% 		\State $states_{key} \leftarrow \sall$
% 		\EndProcedure
% 		%		\State
% 		\Procedure{insert}{key, value}
% 		\If {$states_{key} = \sall$}
% 		\State $states_{key} \leftarrow \sset$
% 		\State $values_{key} \leftarrow \textbf{eval} \; value $
% 		\ElsIf{$states_{key} = \sset$}
% 		\State $values_{key} \leftarrow values_{key} + \textbf{eval} \; value$
% 		\EndIf
% 		\EndProcedure
% 		%		\State
% 		\Procedure{remove}{key}
% 		\If {$states_{key} = \sset$}
% 		\State $states_{key} \leftarrow \snall$
% 		\State \textbf{return} $values_{key}$
% 		\Else
% 		\State $states_{key} \leftarrow \snall$
% 		\State \textbf{return} $none$
% 		\EndIf
% 		\EndProcedure
% 		\EndClass
% 	\end{algorithmic}
% \end{algorithm}

\begin{algorithm}[t]
	\caption{MSA Masked SpGEVM}
	\label{algo:msa_masked_spgevm}
	\begin{algorithmic}[1]
	\Require Sparse row vectors $\vm$, $\vu$, and a sparse matrix $\mB$ 
	\Ensure Sparse row vector $\vvv$
%		\State set vector $v$ to $\emptyset$
		\State $accum \leftarrow \textproc{init}(ncols(\mB))$

		\ForEach{nonzero $\vm_{j}$ in $\vm$}
		\State $accum.\textproc{setAllowed}(j)$
		\EndFor
		
		\ForEach{nonzero $\vu_{k}$ in $\vu$}
		\ForEach{nonzero $\mB_{kj}$ in row $\mB_{k*}$}
		\State $accum.\textproc{insert}(j, \lambda \rightarrow \vu_{k} \mB_{kj})$
		\EndFor
		\EndFor
		
		\ForEach{nonzero $\vm_{j}$ in $\vm$}
		\State $value = accum.\textproc{remove}(j)$
		\If {$value \neq none$}
		\State $\vvv_{j} \leftarrow value$
		\EndIf
		\EndFor
	\end{algorithmic}
\end{algorithm}

\subsection{Hash Accumulator}
\label{sec:acc_hash}

In practice, the arrays in the MSA accumulator are too large to fit in L1 cache, even though they usually have only a few nonzeros, so 
indexing an element of these arrays usually incurs a cache miss in the MSA algorithm ~\cite{Patwary2015}.
To overcome this issue, we utilize a hash map instead of dense arrays for storing values and states, reducing cache misses  but increasing access overhead.
To reduce the  hash  accesses, we store both the accumulated value and its state as a pair in one single hash map, using open addressing with linear probing, no resizing support since we know it will have $\dnnz(\vm)$ values, and a load factor of $0.25$ to reduce collisions.
%\Aydin{there are a couple of other hash accumulators in the literature, cite them and mention differences if any: \cite{deveci2018multithreaded, nagasaka2019performance} }
Others~\cite{nagasaka2019performance,deveci2018multithreaded} use similar hash accumulators for plain SpGEMM.

The time complexity of SpGEVM is $O(\dnnz(\vm) + \flops(\vu \mB))$ operations, since initialization does not depend on $\ncols(\vvv)$ but on the number of nonzeros in mask $\vm$.
Hash accumulator has a smaller memory footprint than MSA, but accessing individual values requires computing the hash.

\subsection{Mask Compressed Accumulator (MCA)}
\label{sec:acc_mca}
\begin{figure}[]
	\centering
	\includegraphics[width=0.35\textwidth]{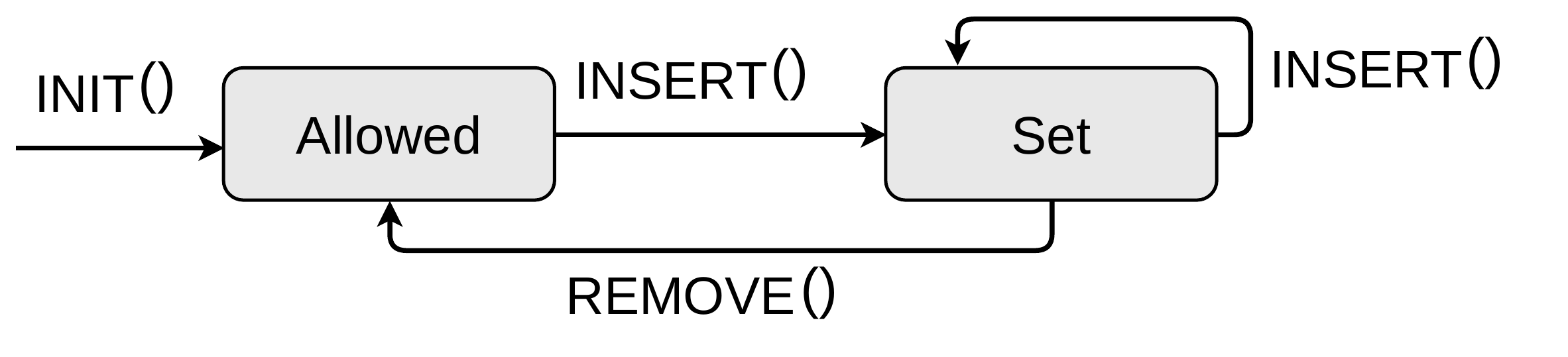}
	\caption{MCA states automaton}
	%\vspace{-.1in}
	\label{fig:mca-states}
\end{figure}

\begin{figure*}[ht]
	%	\centering
	\begin{subfigure}[b]{0.195\textwidth}
		\centering
		\includegraphics[width=\textwidth]{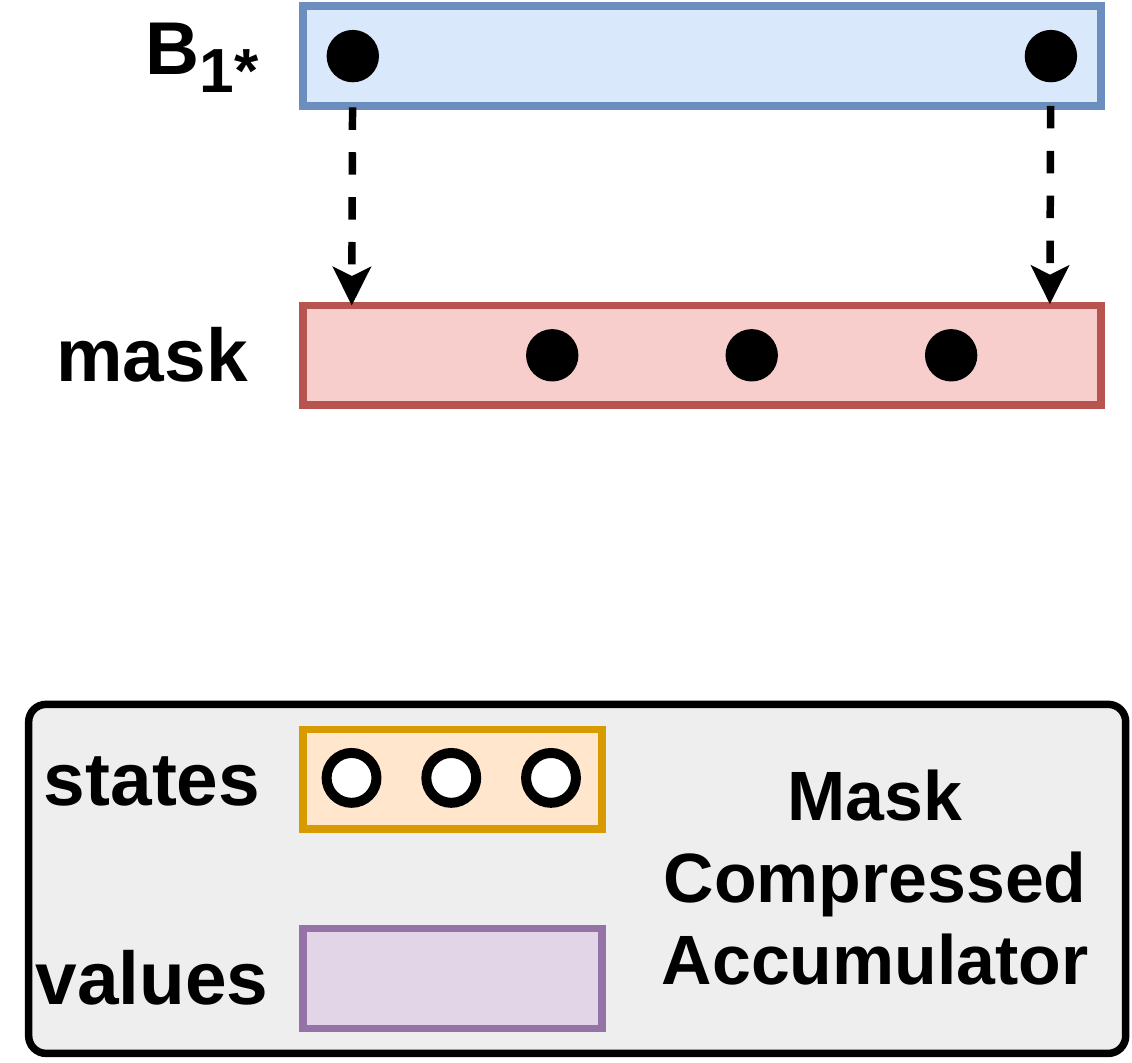}
		\caption{$\textrm{MCA} = \vu_1 \times \mB_{1*}$}
		\label{fig:mca-alg-1}
	\end{subfigure}
	\begin{subfigure}[b]{0.195\textwidth}
		\centering
		\includegraphics[width=\textwidth]{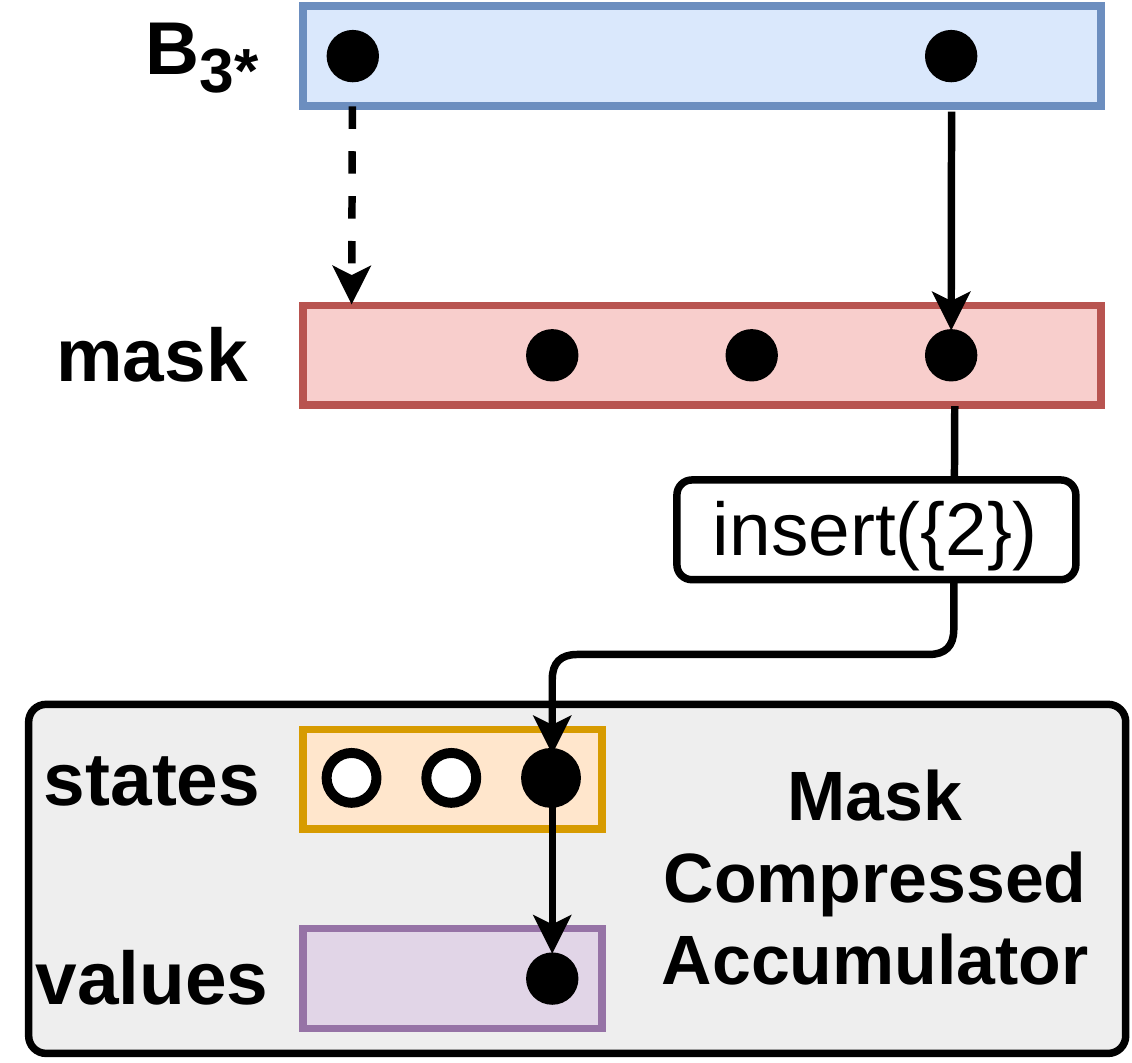}
		\caption{$\textrm{MCA} += \vu_3 \times \mB_{3*}$}
		\label{fig:mca-alg-2}
	\end{subfigure}
	\begin{subfigure}[b]{0.195\textwidth}
		\centering
		\includegraphics[width=\textwidth]{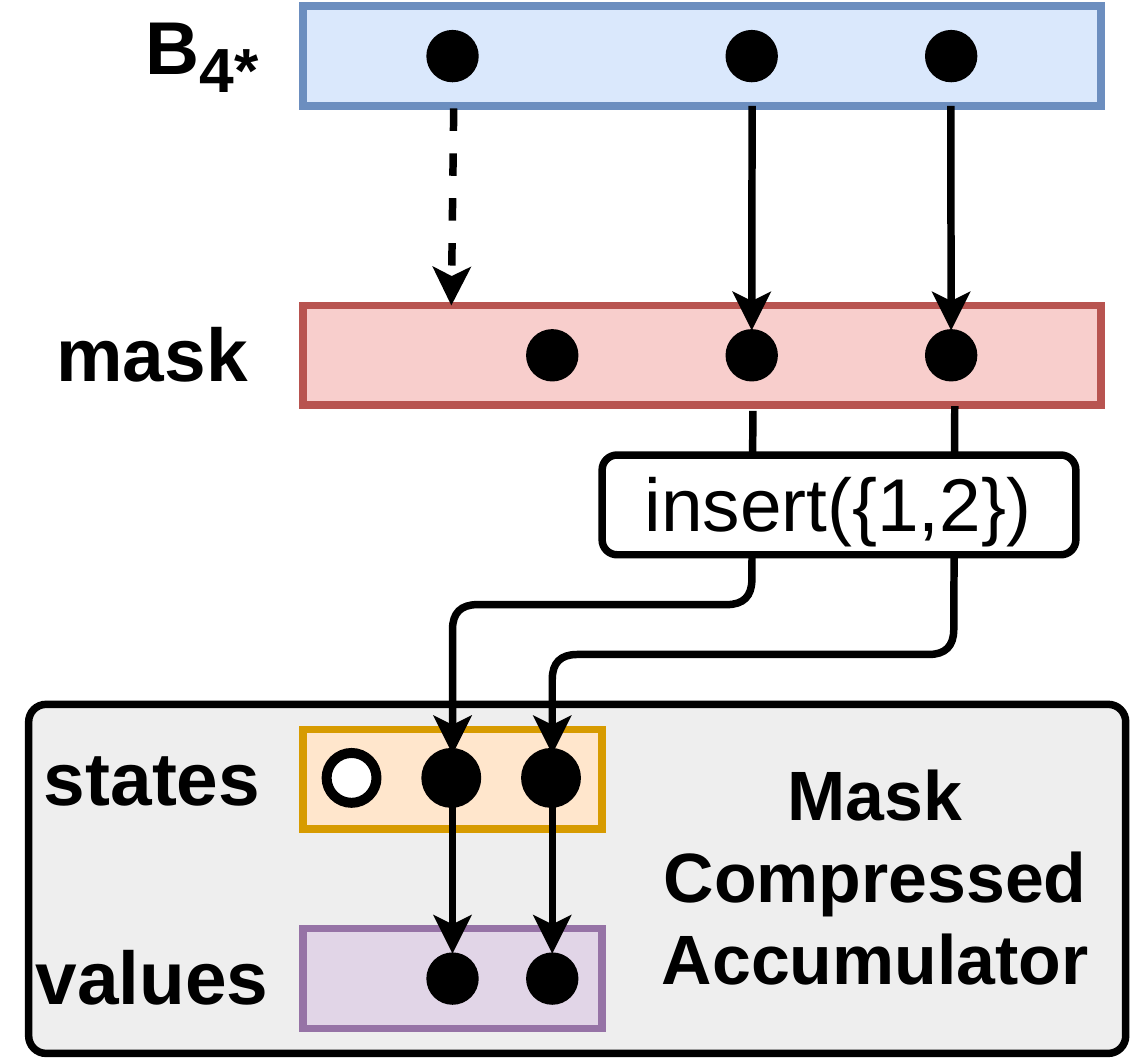}
		\caption{$\textrm{MCA} += \vu_4 \times \mB_{4*}$}
		\label{fig:mca-alg-3}
	\end{subfigure}
	\begin{subfigure}[b]{0.195\textwidth}
		\centering
		\includegraphics[width=\textwidth]{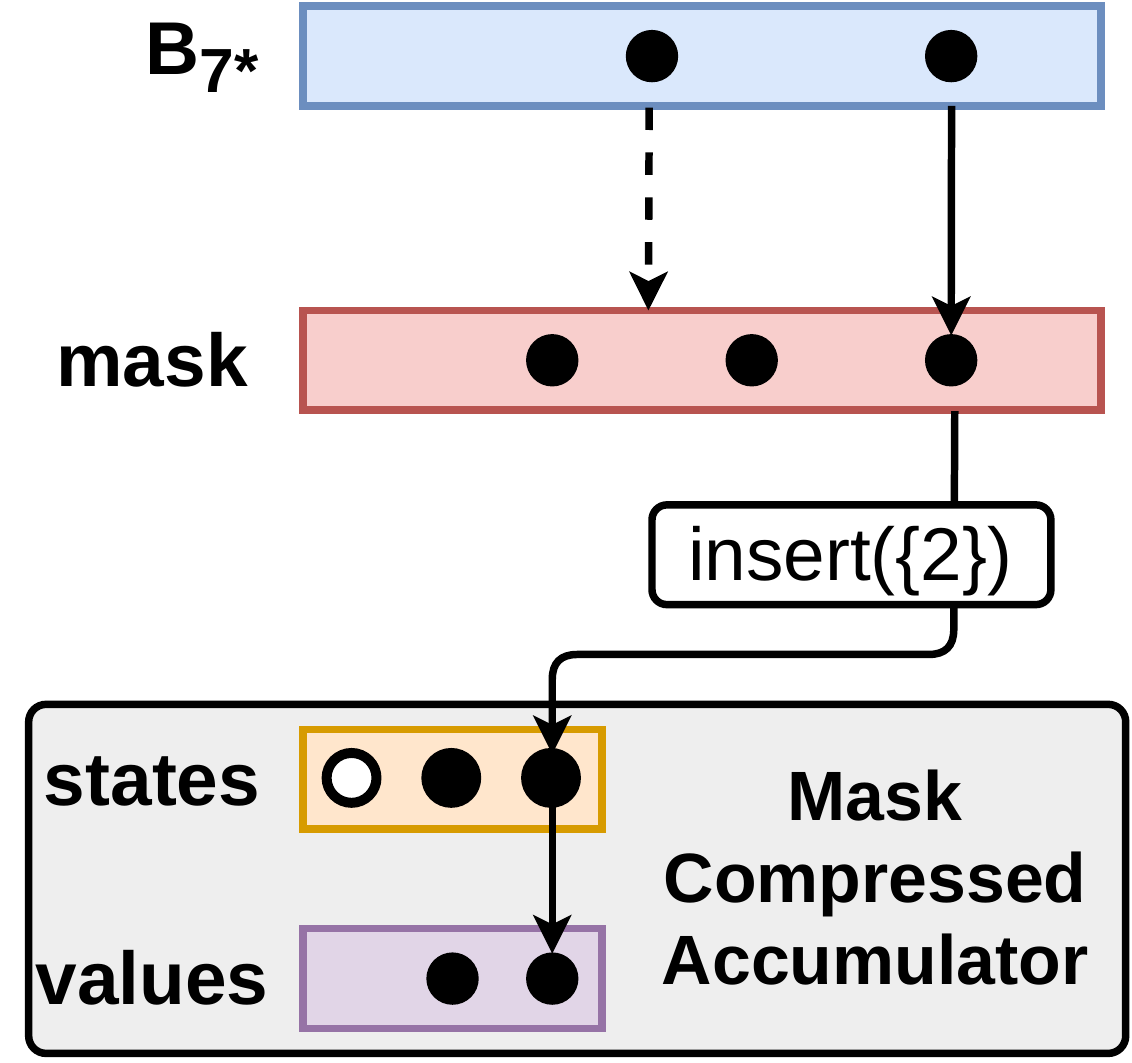}
		\caption{$\textrm{MCA} += \vu_7 \times \mB_{7*}$}
		\label{fig:mca-alg-4}
	\end{subfigure}
	\begin{subfigure}[b]{0.195\textwidth}
		\centering
		\includegraphics[width=\textwidth]{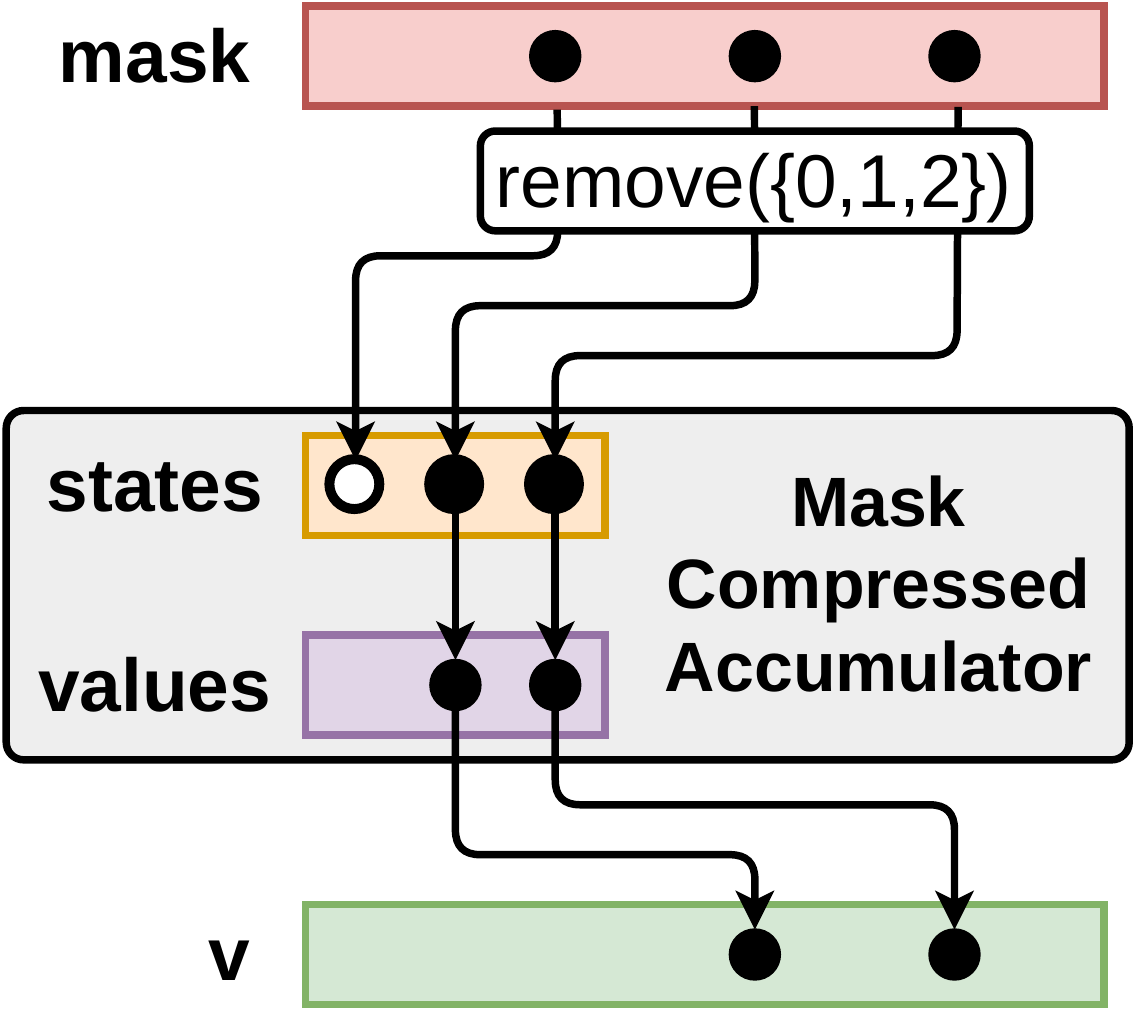}
		\caption{Moving the result to $\vvv$}
		\label{fig:mca-alg-5}
	\end{subfigure}
	\caption{Masked SpGEVM algorithm using Mask Compressed Accumulator (MCA)}.
	%\vspace{-15pt}
	\label{fig:mca}
\end{figure*}

Mask Compressed Accumulator (MCA) algorithm is based on the observation that the number of elements in the accumulator cannot be greater than the number of nonzeros in mask $\vm$.
The MCA accumulator uses size $nnz(\vm)$  for the $values$ and $states$ arrays.
The previously described MSA and Hash algorithms use column indices to index the accumulator. 
However, column indices cannot be used to index the MCA accumulator because $\ncols(\mB)$ is greater than or equal to $\dnnz(\vm)$, i.e., the length of the arrays in the MCA accumulator. 
%
% \Aydin{isn't hash accumulator initialized by the mask element doing something similar here? we map the range $1\ldots n$ to a smaller one defined by the elements in the mask?} \Srdjan{Yes, it is similar. Hash uses a hash function to create the mapping, MCA finds the intersection to create the mapping.}
Therefore, we need another way to index the MCA accumulator.
Since the indices should be in range $[0, \dnnz(\vm))$, when we have a nonzero element in mask with index $j$, and when $\vu_{k} \neq 0\ \mathrm{and}\ \mB_{kj}\neq 0$, we can use the number of nonzero elements in $\vm_j$ with column index smaller than $j$.
The MCA accumulator needs only two states \sall{} and \sset{} because by relying on the mask elements, it readily ensures no key can be in \snall{} state.
Figure ~\ref{fig:mca-states} shows MCA state automaton algorithm and Figure~\ref{fig:mca} shows masked SpGEVM algorithm that generates output vector $\vvv$ using MCA. 

For sorted input, the index for the MCA accumulator can be calculated quickly. %in $O(1)$ time.
%The algorithm starts by finding intersections between column ids of $M_{i*}$ and column ids of elements in rows $\{b_{k*} | a_{ik} \neq 0\}$.
%$Then for each element with column id $k$ from each intersection, the algorithm updates elements in $values$ and $states$ arrays with index $|\{l\ |\ m_{kl} \neq 0\ \mathrm{and}\ l < j\}|$.
For each nonzero in $\vu$ the algorithm iterates through $\vm$ at most once, and it accesses the entries in $\mB_{k*}$ once for which $\vu_{k} \neq 0$ and $\mB_{kj} \neq 0$.
Hence, the algorithm takes $O(\dnnz(\vu) \cdot \dnnz(\vm) + \flops(\vu \mB))$ time.
%

% \begin{algorithm}[t]
% 	\caption{Mask Compressed Accumulator (MCA)}
% 	\label{algo:mca_accumulator}
% 	\begin{algorithmic}[1]
% 		\Class{MaskCompressedAccumulator}
		
% 		\Procedure{init}{maxValues}
% 		\State allocate dense arrays for states and values
% 		\State $states_* \leftarrow \sall$
% 		\EndProcedure
% %		\State

% 		\Procedure{setAllowed}{key}
% 		\State $noop$ 
% 		\EndProcedure
% %		\State
% 		\Procedure{insert}{key, value}
% 		\If {$states_{key} = \sall$}
% 		\State $states_{key} \leftarrow \sset$
% 		\State $values_{key} \leftarrow value$
% 		\Else
% 		\State $values_{key} \leftarrow values_{key} + value$
% 		\EndIf
% 		\EndProcedure
% %		\State
% 		\Procedure{remove}{key}
% 		\If {$states_{key} = \sset$}
% 		\State $states_{key} \leftarrow \sall$
% 		\State \textbf{return} $values_{key}$
% 		\Else
% 		\State \textbf{return} $none$
% 		\EndIf
% 		\EndProcedure
% 		\EndClass
% 	\end{algorithmic}
% \end{algorithm}

\begin{algorithm}[t]
	\caption{MCA Masked SpGEVM}
	\label{algo:mca_masked_spgevm}
	\begin{algorithmic}[1]
	\Ensure Sparse row vectors $\vm$, $\vu$, and a sparse matrix $\mB$ 
	\Require Sparse row vector $\vvv$
%		\State set vector $v$ to $\emptyset$
		\State $accum \leftarrow \textproc{init}(nnz(\vm))$
		
		\ForEach{nonzero $\vu_{k}$ in $\vu$}
%		Find intersection between $m$ and $b_k$
		\State $rowIter \leftarrow MakeIterator(\mB_{k*})$
		\ForEach{$(idx, \vm_j)$ in $Enumerate(\vm)$}
		\While{$rowIter \land rowIter.colId < j$}
		\State $rowIter \leftarrow Next(rowIter)$
		\EndWhile
		
		\If {$rowIter \land rowIter.colId = j$}
		\State $accum.\textproc{insert}(idx, \vu_{k}\mB_{kj})$		
		\EndIf
		\EndFor
		\EndFor
		
		\ForEach{$(idx, \vm_{j})$ in $Enumerate(\vm)$}
		\State $value = accum.\textproc{remove}(idx)$
		\If {$value \neq none$}
		\State $\vvv_{j} \leftarrow value$
		\EndIf
		\EndFor
	\end{algorithmic}
\end{algorithm}

%\begin{figure}[ht]
%%	\centering
%	\begin{subfigure}[b]{0.155\textwidth}
%		\centering
%		\includegraphics[width=\textwidth]{figures/mca-1-5.pdf}
%		\caption{Adding $\vu_1 \times \mB_{1*}$ to MCA}
%		\label{fig:mca-alg-1}
%	\end{subfigure}
%	\begin{subfigure}[b]{0.155\textwidth}
%		\centering
%		\includegraphics[width=\textwidth]{figures/mca-2-5.pdf}
%		\caption{Adding $\vu_3 \times \mB_{3*}$ to MCA}
%		\label{fig:mca-alg-2}
%	\end{subfigure}
%	\begin{subfigure}[b]{0.155\textwidth}
%		\centering
%		\includegraphics[width=\textwidth]{figures/mca-3-5.pdf}
%		\caption{Adding $\vu_4 \times \mB_{4*}$ to MCA}
%		\label{fig:mca-alg-3}
%	\end{subfigure}
%	\begin{subfigure}[b]{0.155\textwidth}
%		\centering
%		\includegraphics[width=\textwidth]{figures/mca-4-5.pdf}
%		\caption{Adding $\vu_7 \times \mB_{7*}$ to MCA}
%		\label{fig:mca-alg-4}
%	\end{subfigure}
%	\begin{subfigure}[b]{0.155\textwidth}
%		\centering
%		\includegraphics[width=\textwidth]{figures/mca-5-5.pdf}
%		\caption{Moving the result to $\vvv$}
%		\label{fig:mca-alg-5}
%	\end{subfigure}
%	\caption{}
%	\label{fig:mca}
%\end{figure}

\subsection{Masked Heap SpGEVM Algorithm}
\label{sec:acc_heap}
In this section we describe a Masked SpGEVM algorithm based on column-column Heap algorithm developed by Buluç and Gilbert~\cite{buluc2008representation}. 
Like the base algorithm, our heap algorithm requires that the indices in mask $\vm$ and column indices in matrix $\mB$ are sorted.
To compute the output vector $\vvv$, our algorithm uses a min-heap (or a priority queue).
The heap initially  contains $\dnnz(\vu)$ row iterators that point to the first element of rows $\{\mB_{k*}$ | $\vu_k \neq  0\}$.
The iterators in the heap are ordered based the column index of the element they point to.
By popping the top iterator from the heap, incrementing it, and pushing the incremented iterator back to the heap, we can iterate through set $\mathbf{S} = \{\mB_{kj} | \vu_k \neq  0\}$ in sorted order without having to construct the sorted set in memory.
This is similar to the multi-way merge from \cite{10.5555/280635}.
Since the indices in mask $\vm$ are also sorted, we can easily find the intersection between the mask and set $\mathbf{S}$ by performing a 2-way merge.
For all elements $\mB_{kj}$ that belong to the intersection, the algorithm calculates $\vu_k \mB_{kj}$ products and inserts the result to the output.
If the last inserted product has the same column index as the product currently being inserted, the result of the current product is added to the last product. 
Otherwise, a new entry is added to the output.
Algorithm~\ref{algo:heap_masked_spgevm} shows the pseudo-code for the masked heap SpGEVM algorithm and Algorithm~\ref{algo:insert_heap} shows the insert procedure for the heap.

%% PAR complexity
The heap iterates through the mask, and pops $\flops(\vu \mB)$ elements from the heap.
The complexity of the algorithm is $O(\dnnz(\vm) + \log_2\dnnz(\vu) \cdot \flops(\vu \mB))$, since the maximum heap size is $\dnnz(\vu)$.
Compared to MSA and Masked Hash algorithm, heap algorithm has smaller memory footprint but greater asymptotic complexity due to the logarithmic factor.
To alleviate this, we can exploit the fact that only the elements in the intersection $\vm \cap \mathbf{S}$ will be used to form the output and we can check whether the element will be a part of the intersection before we push it to the heap.
Such a check might require us to iterate through the whole mask.
For this reason, we configure the algorithm to inspect only a portion of the mask $\vm$ before pushing an element to the heap by using the NInspect parameter (Algorithm~\ref{algo:insert_heap}), which controls the length of the portion of the mask to be checked.
If this parameter is 0, the algorithm is identical to the algorithm described earlier.
If it is 1, the algorithm inspects only the current mask element and has complexity $O(\dnnz(\vm) +  (\alpha + (1 - \alpha) \cdot \log_2\dnnz(\vu)) \cdot \flops(\vu \mB))$, where $\alpha$  depends on the input and $\alpha \in [0,1]$. 
In our experiments we test 1 and $\infty$ for the NInspect parameter.

%% PAR complement
For mask complement, the only difference is that instead of computing $\vu_k \mB_{kj}$ products for the elements in intersection $\vm \cap \mathbf{S}$, we compute the products for the elements in set difference $\mathbf{S} \setminus \vm$. 
When mask is complemented, NInspect parameter is always 0.

\begin{algorithm}[t]
	\caption{Heap Masked SpGEVM}
	\label{algo:heap_masked_spgevm}
	\begin{algorithmic}[1]
	\Require Sparse row vectors $\vm$, $\vu$, and a sparse matrix $\mB$ 
	\Ensure Sparse row vector $\vvv$
%		\State set vector $v$ to $\emptyset$

		\State $mIter \leftarrow MakeIterator(\vm)$

		\ForEach{$\vu_{k}$ in $\vu$}
		\State $\textproc{insert}(PQ, MakeIterator(\mB_{k*}), mIter)$
		\EndFor
		
%		\State
		
		\State $prevKey \leftarrow none$
		\While{$\neg PQ.isEmpty()$}
		
		\State $minIter \leftarrow PQ.popMin()$
		\While{$mIter \land mIter.colId < minIter.colId$}
		\State $mIter \leftarrow Next(mIter)$
		\EndWhile
		\If{$\neg mIter$}
		\State \Break
		\EndIf
		
%		\State
		
		\If{$mIter.colId = minIter.colId$}
		\State $k \leftarrow minIter.rowId$
		\State $j \leftarrow minIter.colId$
		\If{$prevKey = minIter.colId$}
		\State $\vvv_j \leftarrow \vvv_j + \vu_{k} \mB_{kj}$
		\Comment{$\vvv_j$ is the last value in $\vvv$}
		\Else
		\State $prevKey \leftarrow minIter.colId$
		\State $\vvv_j \leftarrow \vu_{k} \mB_{kj}$
		\Comment{$\vvv_j$ is the last value in $\vvv$}
		\EndIf
		\EndIf
		
		\State $\textproc{insert}(PQ, Next(minIter), mIter)$
		
		\EndWhile		

	\end{algorithmic}
\end{algorithm}

\begin{algorithm}[t]
	\caption{Insert procedure for Heap Masked SpGEVM}
	\label{algo:insert_heap}
	\begin{algorithmic}[1]
%		\Procedure{Insert1}{PQ, rowIter, mIter}
%		\If{$IsValid(rowIter)$}
%		\State $PQ.insert(rowIter)$;
%		\EndIf
%		\EndProcedure	
%
%		\Procedure{Insert2}{PQ, rowIter, mIter}
%		\While{$IsValid(rowIter)$}
%		\If{$rowIter.colId >= mIter.colId$}
%		\State $PQ.insert(rowIter)$
%		\State \Return
%		\Else
%		\State $rowIter \leftarrow Next(rowIter)$
%		\EndIf
%		\EndWhile
%		\EndProcedure
%		
%		\Procedure{Insert3}{PQ, rowIter, mIter}
%		
%		\While{$IsValid(rowIter) \land IsValid(mIter)$}
%		\If{$rowIter.colId = mIter.colId$}
%		\State $PQ.insert(rowIter)$;
%		\State \Return
%		\ElsIf{$rowIter.colId < mIter.colId$}
%		\State $rowIter \leftarrow next(rowIter)$
%		\Else
%		\State $mIter \leftarrow next(mIter)$		
%		\EndIf
%		\EndWhile
%		\EndProcedure
		
		\Procedure{Insert}{PQ, rowIter, mIter, NInspect}
		\If{$\neg IsValid(rowIter)$}
		\State \Return
		\EndIf
		
		\If {$NInspect = 0$}
		\State $PQ.insert(rowIter)$;
		\State \Return
		\EndIf
		
		\State $toInspect \leftarrow NInspect$
		\While{$IsValid(rowIter) \land IsValid(mIter)$}
		\If{$rowIter.colId = mIter.colId$}
		\State $PQ.insert(rowIter)$;
		\State \Return
		\ElsIf{$rowIter.colId < mIter.colId$}
		\State $rowIter \leftarrow Next(rowIter)$
		\Else
		\State $mIter \leftarrow Next(mIter)$
		\State $toInspect \leftarrow toInspect - 1$
		\If{$toInspect = 0$}
		\State $PQ.insert(rowIter)$;
		\State \Return
		\EndIf
		\EndIf
		\EndWhile
		
		\EndProcedure

	\end{algorithmic}
\end{algorithm}
\section{Symbolic and Numeric Phases}
\label{sec:phase}
The size and the pattern of the output matrix in the Masked SpGEMM are not known before the multiplication.
To allocate the needed space and form the output matrix, there are two methods: one-phase and two-phase approaches.

%% one-phase
In the one-phase approach, the Masked SpGEMM is performed all at once,
first allocating temporary memory large enough to store the output matrix, executing the Masked SpGEMM, and then copy the values from the temporary memory to the output matrix.
This is often deemed inefficient in plain SpGEMM, especially when the compression factor is large. However, mask can provide a good initial approximation for the size of the output matrix, making one-phase approaches  more viable for Masked SpGEMM.
%Since a large portion of the output matrix may be discarded by the mask, copying values from the temporary memory might be cheaper than executing the symbolic phase.

%% two-phase
In the two-phase approach, we first execute a symbolic multiplication that only inspects the row and column indices from the inputs to computes the number of nonzeros in the output matrix, then allocate memory for the output matrix and execute the actual multiplication.
The former is known as the symbolic and the latter is known as the numeric phase.
%The part of the algorithm where we count the number of nonzeros in the output is known as symbolic phase, and the part of the algorithm where we perform the actual multiplication is known as numeric phase.
%

%%
The trade-off between these two approaches is in memory footprint vs. amount of computation.
The two-phase approach reduces the memory footprint  at the expense of increased computation.
%% explain the tradeoff for running 1-phase and 2-phase
We evaluate both approaches for the algorithms described in our work since the addition of the mask to the multiplication has the potential to alter the balance between these trade-offs for plain SpGEMM.

% \input{text/analysis}

% \section{Hybrid algorithm}
% We can choose the best algorithm for each row of the output. For that, we use the number of (unmasked) sparse flops, nonzeros on each row of the mask as well as the inputs, as proxies for choosing the best algorithm. 

% Our algorithms do not have a symbolic phase but they do perform a scan of the inputs for sparse flops estimation. 

\section{Experimental Setup}
\label{sec:setup}
%% PAR matrices
We conduct our experiments on both synthetic and real-world graphs.
The synthetic graphs are preferred for controlled experiments in which we vary degree or size of the graphs and investigate their effects on various performance metrics.
For the synthetic graphs, we utilize \erdosrenyi{} graphs as well as graphs generated with R-MAT generator~\cite{Chakrabarti2004}, with parameters identical to those used in the Graph500 benchmark~\cite{murphy2010introducing}.
For real-world graphs, we use the same set of 26 graphs that Nagasaka et al.~\cite{nagasaka2019performance} (Table 2 of the referenced paper) used, which are all from SuiteSparse Matrix Collection~\cite{Davis2011}. 
Since we only plot performance profiles~\cite{Dolan2002} where we report on the fraction of input cases as a function of the relative runtime of each algorithm, we do not repeat the graph properties here. Their input nonzeros range from $350$K to $100$M.   %as  whose various properties are listed in Table~\ref{tab:florida_mat}.

%% PAR schemes

%% PAR system/code/compilers
We conduct our experiments on two different systems: {\bf Haswell} with Intel Xeon E5-2698 processors (two sockets per node, 2.3 GHz, 32 total cores, 128GB) and {\bf KNL} with Intel Xeon Phi 7250 processors (one socket per node, 1.4 GHz, 68 cores, 96 GB).
All algorithms are implemented in C++, compiled with gcc v10.1 with -O3 flag.
Threads are pinned to cores using GOMP\_CPU\_AFFINITY.
As a baseline, we use SuiteSparse:GraphBLAS version 5.1.4 compiled with the same parameters as above.
With the exception of scaling experiments, we use 32 threads on Haswell and 68 threads on KNL in all experiments.

%% PAR benchmarks
We benchmark three different applications on real-world graphs: (i) Triangle Counting, (ii) $k$-truss, and (iii) Betweenness Centrality.
Triangle Counting computes the total number of triangles in a graph, using one Masked SpGEMM operation along with a reduction.
$k$-truss finds the edges that are supported by at least $k$-2 other edges, 
using Masked SpGEMM in an iterative manner where the the graph keeps changing due to pruning of some edges.
The Betweenness Centrality measures how central is a node in the graph by computing the ratio of shortest paths that node is on~\cite{Freeman1977}, 
using a multi-source two-stage algorithm~\cite{Brandes2001}.
%
% This algorithm consists of a forward and backward stage in both of which a Masked SpGEMM is performed along with other operations.
%
% The forward stage utilizes a complemented Masked SpGEMM while the 
All these algorithms are implemented within the GraphBLAS specifications, substituting Masked SpGEMM operations with calls to different algorithms investigated in this work to measure their performance.

\section{Experimental Results}
\label{sec:exp}
% \Aydin{Srdjan includes plots 50\% of their existing size; put Haswell/KNL side-by-side, for performance profiles of TC that include 1P vs 2P numbers, only include Haswell and say the KNL numbers are almost identical}

In this section we compare several algorithms on three benchmarks: Triangle Counting, $k$-truss, and Betweenness Centrality.
We evaluated the following schemes:
    {\bf Inner} (pull-based inner-product-parallel algorithm from Section~\ref{sec:pull_algs}), 
    {\bf MSA} (push-based algorithm using masked sparse accumulator from Section~\ref{sec:acc_msa}),
    {\bf Hash} (push-based algorithm using hash accumulator from Section~\ref{sec:acc_hash}),
    {\bf MCA} (push-based algorithm using compressed mask accumulator from Section~\ref{sec:acc_mca}),
    {\bf Heap} (push-based algorithm using a heap accumulator with $NInspect=1$ from  Section~\ref{sec:acc_heap}),
    {\bf HeapDot}  (push-based algorithm using a heap accumulator with $NInspect=\infty$ from Section~\ref{sec:acc_heap}), and 
    two variants of Masked SpGEMM from SuiteSparse:GraphBLAS~\cite{Davis2019} (\SSGB{}) library: {\bf SS:DOT}, a pull-based algorithm  similar to Inner, and {\bf SS:SAXPY}, a push-based algorithm that, depending on the problem, can use SPA-like data structure or a hash table to accumulate values.

%Our algorithms include Inner product (Section\ref{sec:pull_algs}), MSA, Hash, MCA, and Heap (Section \ref{sec:algs}).
%Heap algorithm has two variants, Heap and HeapDot.
%For Heap, NInspect is set to 1, and for HeapDot, NInspect is set to $\infty$.
Each of our algorithms can be executed with and without a symbolic phase, which are respectively indicated with suffixes 2P and 1P.
In total, we evaluate 14 algorithms, 10 of which are proposed in this work, 2 are based on the previous work~\cite{yang2019graphblast}, and 2 of them from \SSGB{} are used as baseline.

% From \SSGB{} we selected DOT and SAXPY algorithms.
% SS:DOT algorithm is a pull based algorithm very similar to Inner product.
% SS:SAXPY algorithm is push based algorithm that, depending on problem, can use SPA-like data structure or Hash to accumulate values.

\subsection{Effect of Input Matrix and Mask Density}
\begin{figure}
	\begin{subfigure}[b]{0.47\textwidth}
		\centering
		\includegraphics[width=\textwidth]{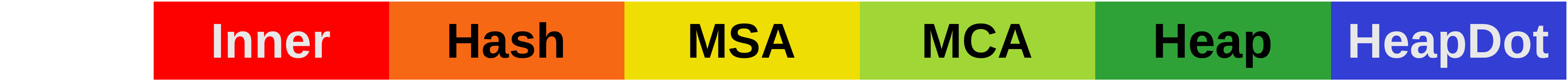}
	\end{subfigure}
	\begin{subfigure}[b]{0.025\textwidth}
		\centering
		\includegraphics[width=\textwidth]{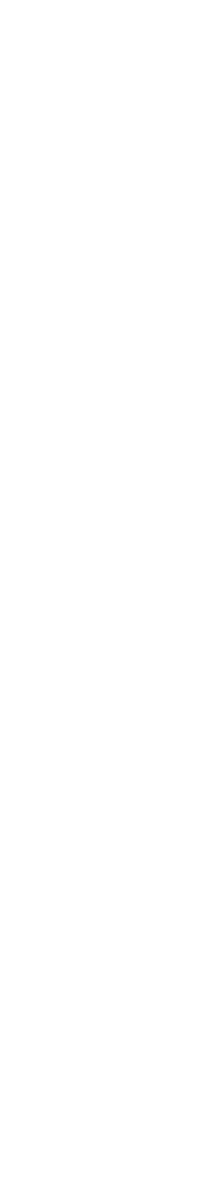}
	\end{subfigure}
    \begin{subfigure}[b]{0.22\textwidth}
		\centering
		\includegraphics[width=\textwidth]{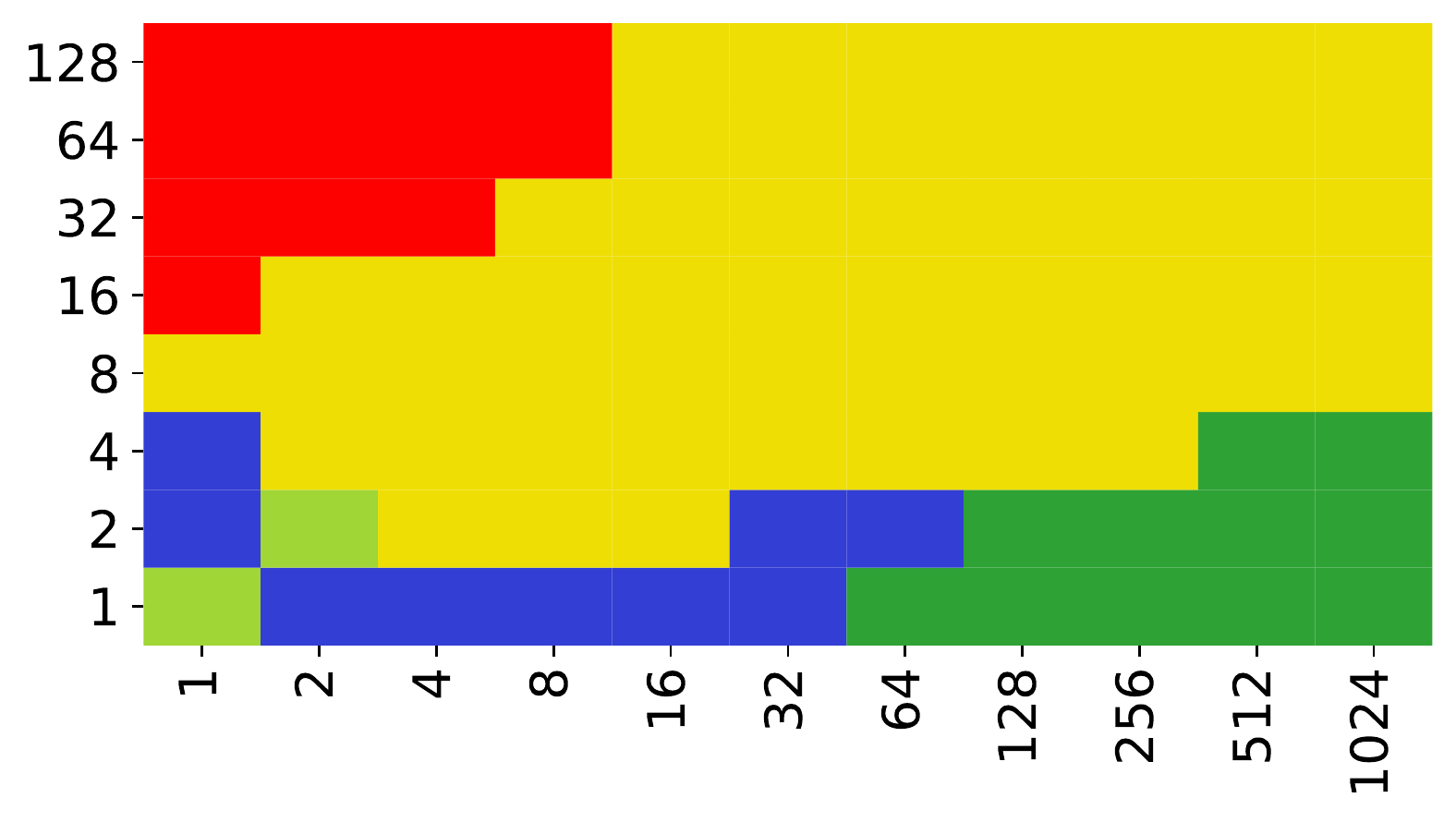}
	%	\vspace{-.25in}
		\caption{dimension = $2^{12} \times 2^{12}$}
	\end{subfigure}
	\begin{subfigure}[b]{0.22\textwidth}
		\centering
		\includegraphics[width=\textwidth]{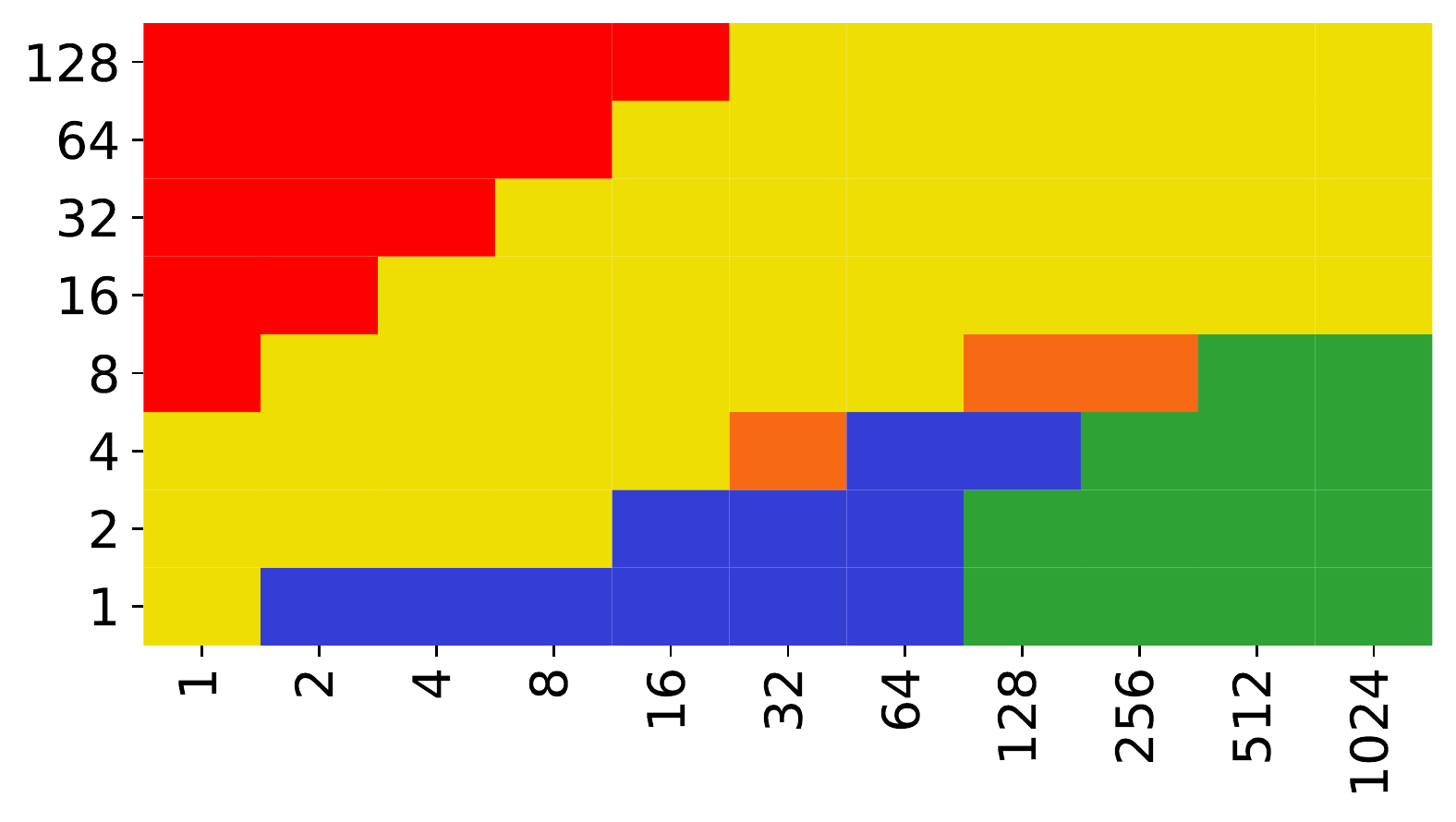}
	%	\vspace{-.25in}
		\caption{dimension = $2^{14} \times 2^{14}$}
	\end{subfigure}
		\begin{subfigure}[b]{0.025\textwidth}
		\centering
		\includegraphics[width=\textwidth]{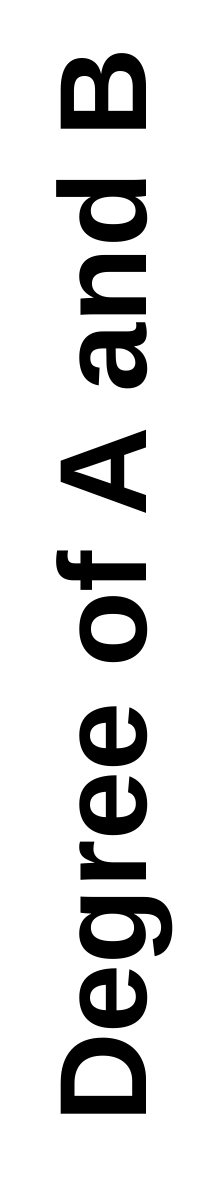}
	\end{subfigure}
	\begin{subfigure}[b]{0.22\textwidth}
		\centering
		\includegraphics[width=\textwidth]{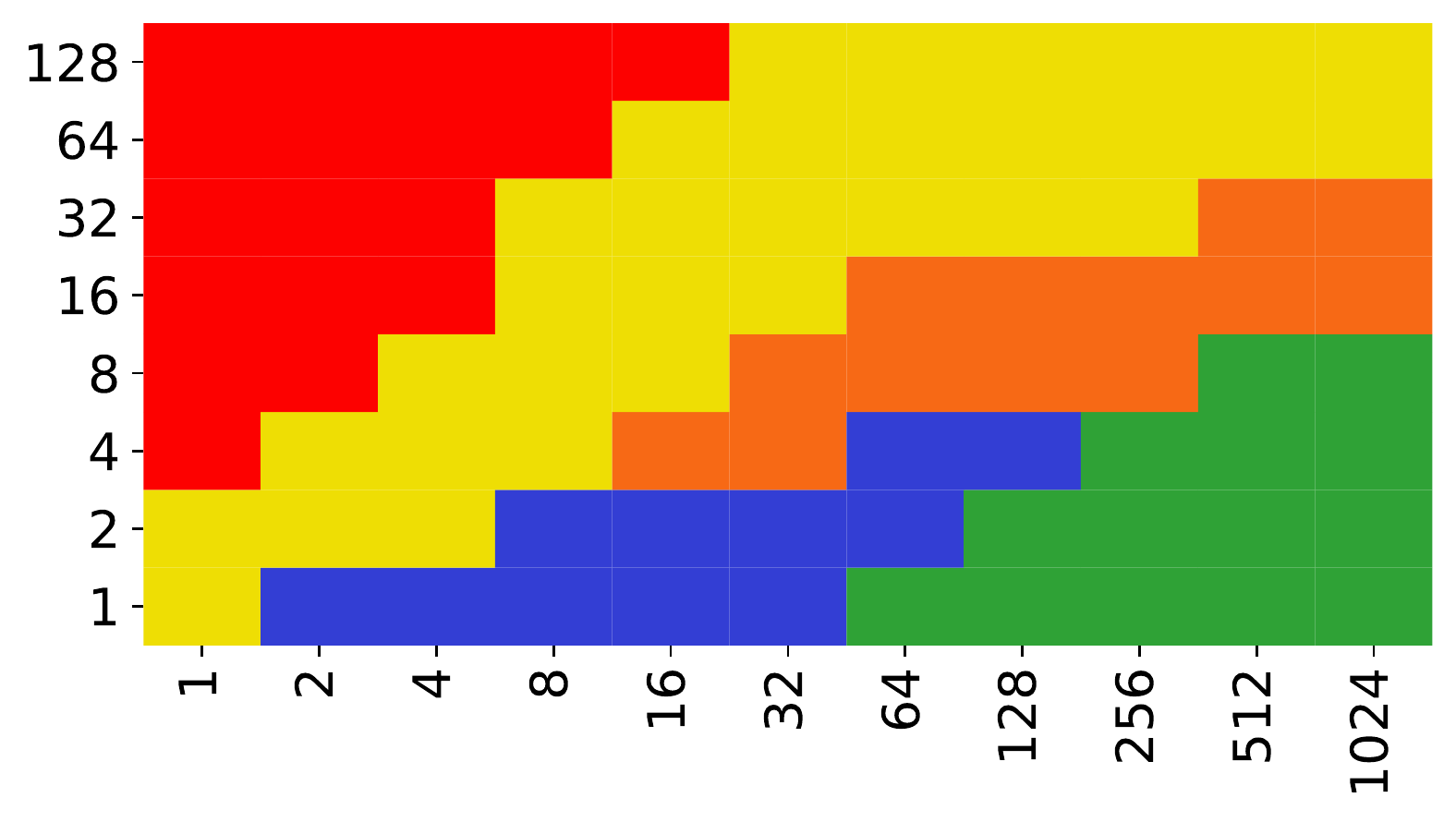}
	%	\vspace{-.25in}
		\caption{dimension = $2^{16} \times 2^{16}$}
	\end{subfigure}
	\begin{subfigure}[b]{0.22\textwidth}
		\centering
		\includegraphics[width=\textwidth]{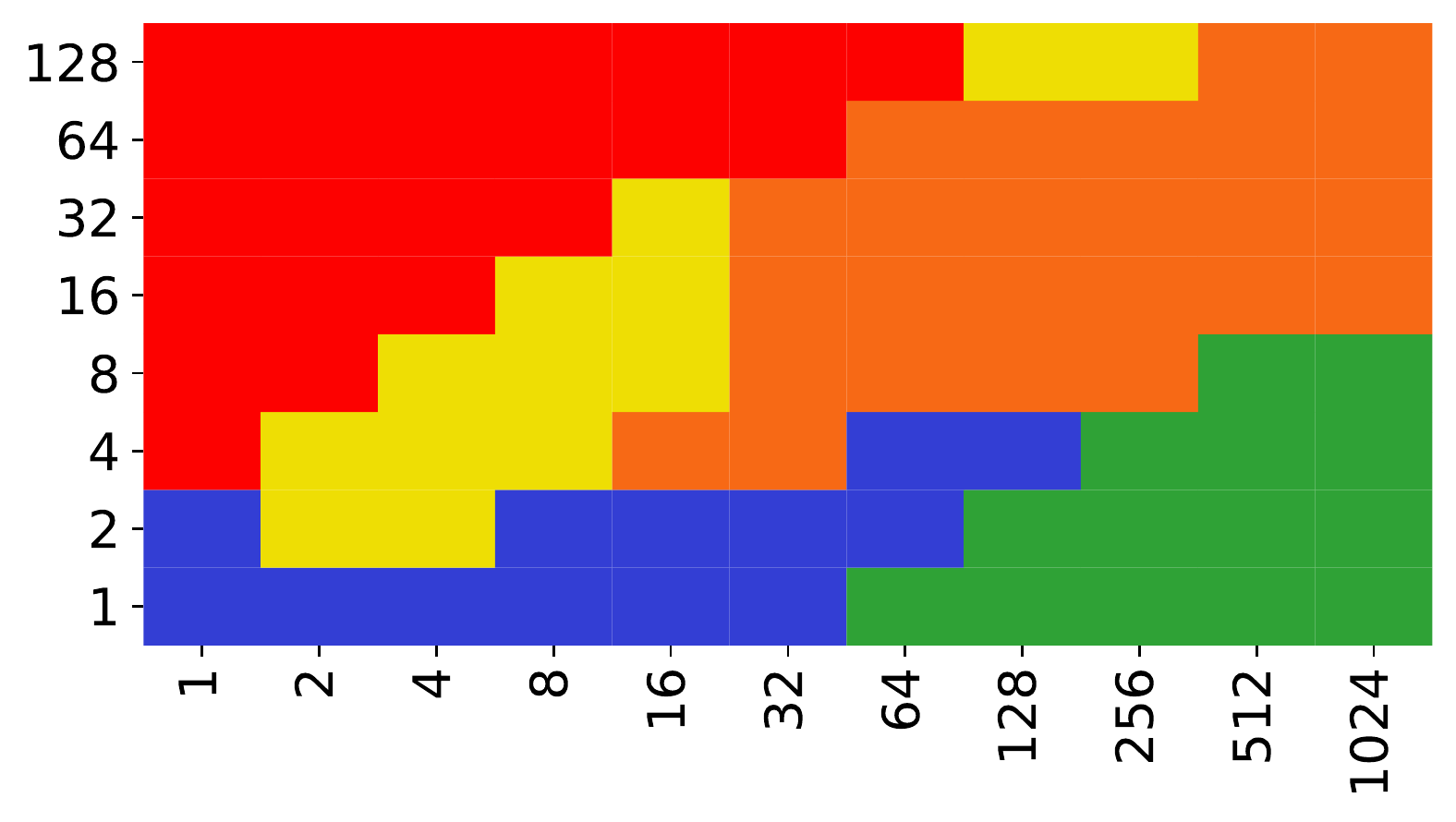}
	%	\vspace{-.25in}
		\caption{dimension = $2^{18} \times 2^{18}$}
	\end{subfigure}
		\begin{subfigure}[b]{0.025\textwidth}
		\centering
		\includegraphics[width=\textwidth]{plots/heatmap/heat-yaxis-empty.pdf}
	\end{subfigure}
	\begin{subfigure}[b]{0.22\textwidth}
		\centering
		\includegraphics[width=\textwidth]{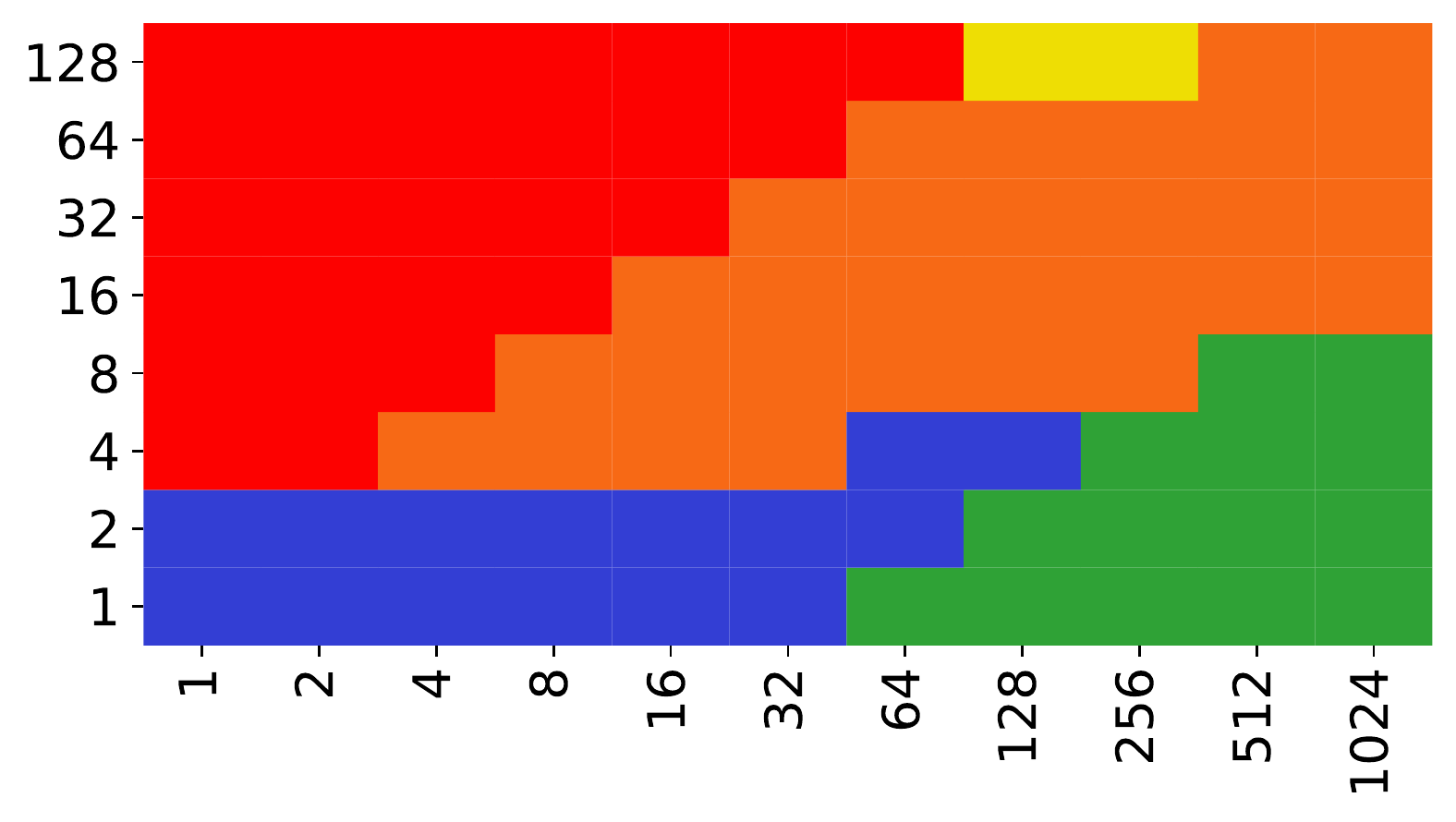}
	%	\vspace{-.25in}
		\caption{dimension = $2^{20} \times 2^{20}$}
	\end{subfigure}
	\begin{subfigure}[b]{0.22\textwidth}
		\centering
		\includegraphics[width=\textwidth]{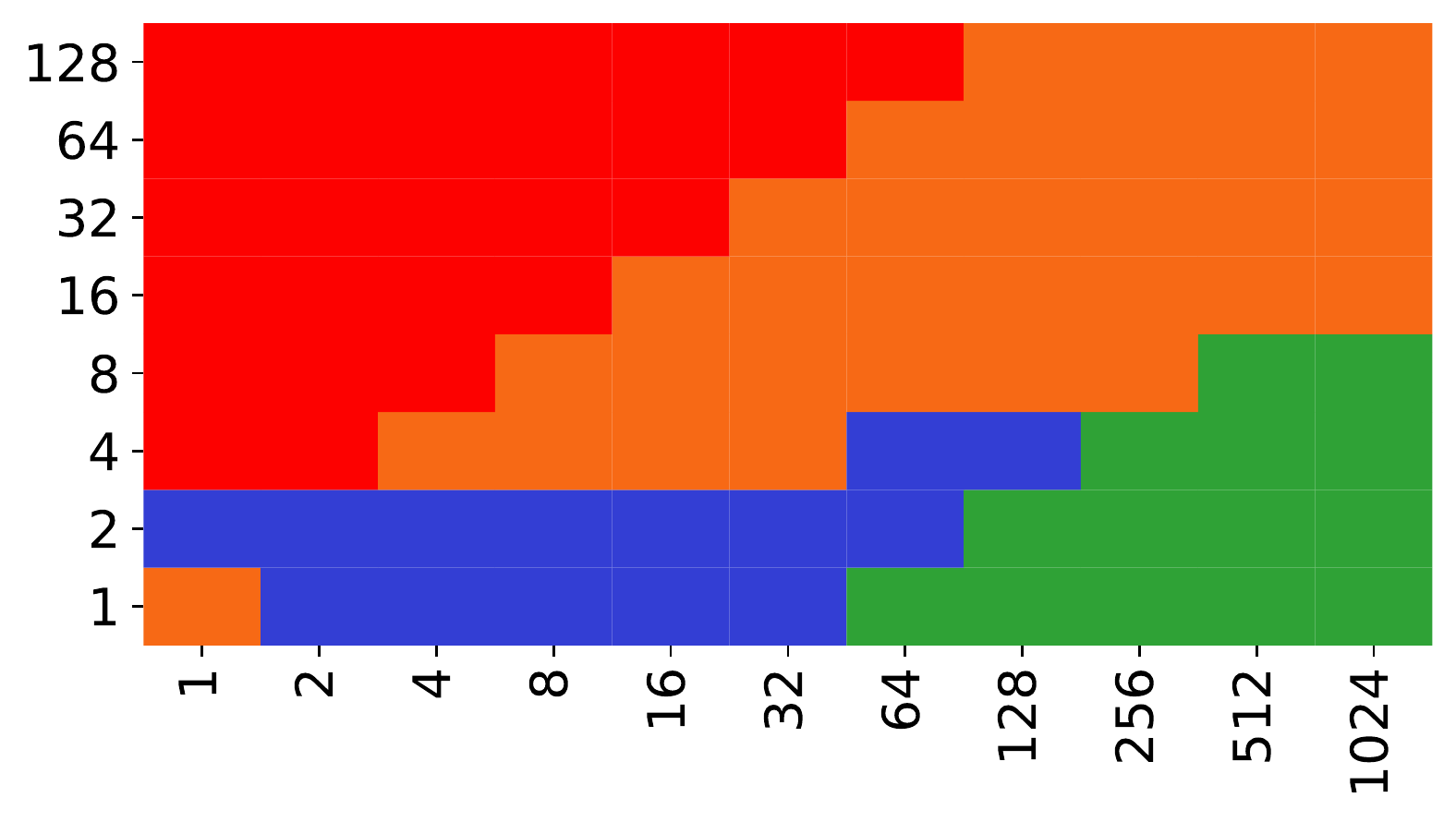}
	%	\vspace{-.25in}
		\caption{dimension = $2^{22} \times 2^{22}$}
	\end{subfigure}
	\begin{subfigure}[b]{0.14\textwidth}
		\centering
		\includegraphics[width=\textwidth]{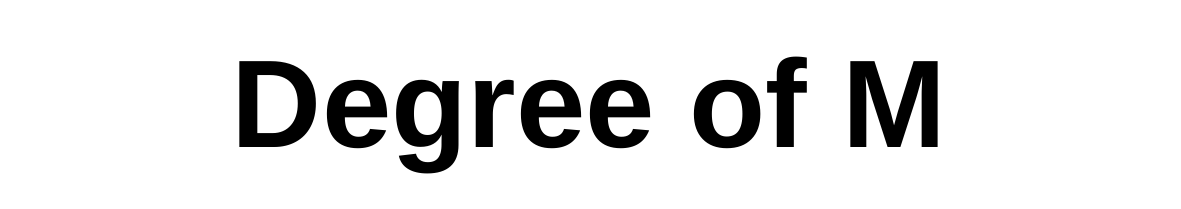}
	\end{subfigure}
    \caption{The best performing schemes with varying input and mask density.}
    \label{fig:heatmap}
\end{figure}

In this section we investigate the performance of our Masked SpGEMM algorithms with changing mask and input matrix density.
These experiments are conducted on Haswell.
Figure~\ref{fig:heatmap} plots the best performing algorithm for multiple different \erdosrenyi{} inputs by varying the degree of the mask in $x$ axis and the degrees of the input matrices in $y$ axis.

When mask is much sparser than $\mA$ and $\mB$, Inner has the best performance, which is because Inner is able to avoid a great amount of unnecessary operations that the other algorithms suffer from.
When $\mA$ and $\mB$ are much sparser than mask, on the other hand, Heap and HeapDot perform the best.
In all other cases where mask and input matrix density are comparable, MSA and Hash show the best performance, with MSA performing better on smaller matrices and Hash on larger ones -- which can be attributed to MSA's worsening cache utilization as the matrices get larger.

\subsection{Triangle Counting}
We next benchmark performance all schemes on Triangle Counting benchmark.
%
%Triangle counting is an important kernel used to characterized graphs\cite{azad2015parallel}.
For optimal performance, vertices in the original graph should be sorted in non-increasing order of their degrees~\cite{lumsdaine2020triangle}.
%in such way that the rows in the resulting matrix are ordered by the number of nonzeros in decreasing order.
After relabeling, the number of triangles is given by as $sum(\mL .* (\mL^2))$, where $\mL$ is the lower triangular matrix.
This method is known to be among the fastest ways to compute Triangle Counting~\cite{Davis2018}.
In our experiments, we only report the Masked SpGEMM execution time.

\begin{figure}
	\centering
	\includegraphics[width=0.45\textwidth]{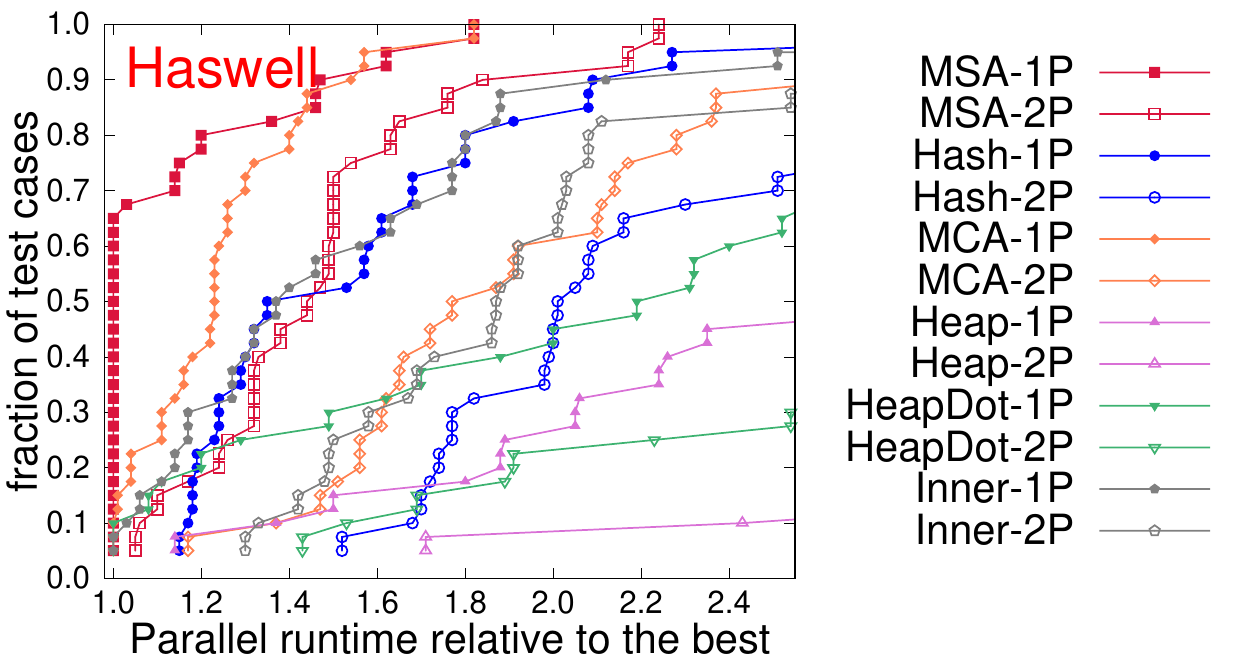}
	\caption{Triangle Counting: our algorithms}
%	\vspace{-.1in}
	\label{fig:tricnt-knl26-haswell-1p2p}
\end{figure}

\noindent
\textbf{Relative Performance of Masked SpGEMM Algorithms.}
Figure~\ref{fig:tricnt-knl26-haswell-1p2p} shows the performance profiles of all our algorithms tested on all real graphs.
In the performance profile plots~\cite{Dolan2002}, a point $(x,y)$ indicates that the scheme for that point is within $x$ factor of the best obtained result in $y$ fraction of the test cases.
The closer a scheme's line is to the $y$ axis, the better is its performance.

In this benchmark, the best performing scheme is MSA-1P, outperforming all other algorithms for 65\% of the test cases, followed by MCA-1P.
These are followed by Inner and Hash schemes, with Heap and HeapDot being the worst.
Observe that the one-phase variant of each algorithm performs better than than its two-phase variant.
We exclude two-phase variants and heap-based schemes from our discussions in this section to keep our plots more readable.
%MSA is followed by MCA, Inner product, and Hash algorithm.
%Heap and HeapDot were generally the slowest algorithms and they are excluded from the other plots.
%The algorithm variants without symbolic phase have better performance compared to their variants with symbolic phase.

\begin{figure}
	\centering
	\includegraphics[width=0.4\textwidth]{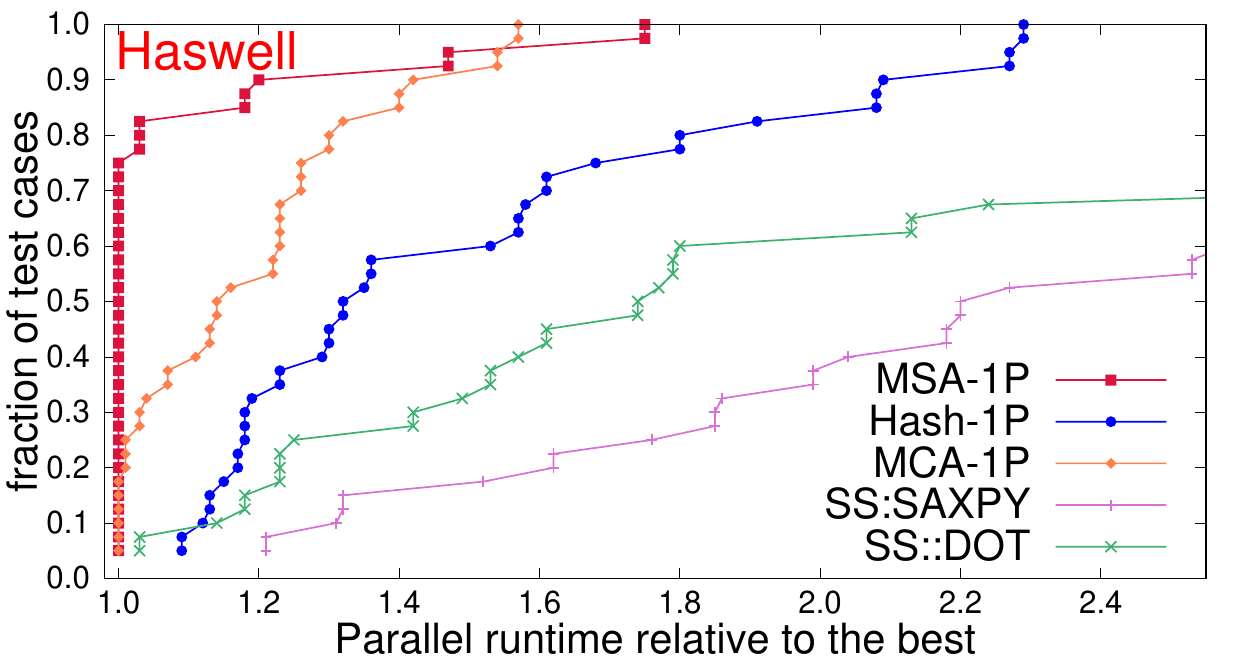}
	\caption{Triangle Counting: our algorithms vs. \SSGB{}.}
%	\vspace{-.1in}
	\label{fig:tricnt-knl26-haswell-summary}
\end{figure}

Figure~\ref{fig:tricnt-knl26-haswell-summary} compares the performance of our three best performing algorithms against the \SSGB{} algorithms.
We can see that all our algorithms outperform \SSGB{} algorithms in almost all cases. Performance profiles are almost identical on KNL and excluded due to space.

\begin{figure}
\begin{subfigure}[b]{0.45\textwidth}
		\centering
		\includegraphics[width=0.8\textwidth]{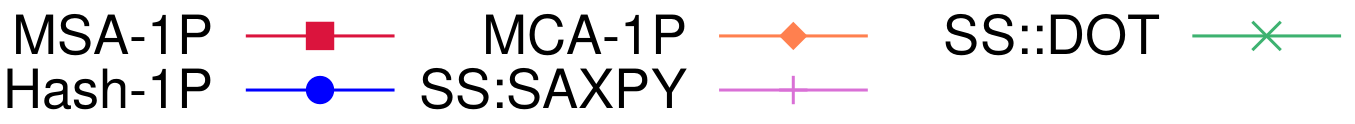}
		\label{fig:tricnt-rmat-knl-flops}
	\end{subfigure}
	\begin{subfigure}[b]{0.5\textwidth}
		\centering
		\includegraphics[width=\textwidth]{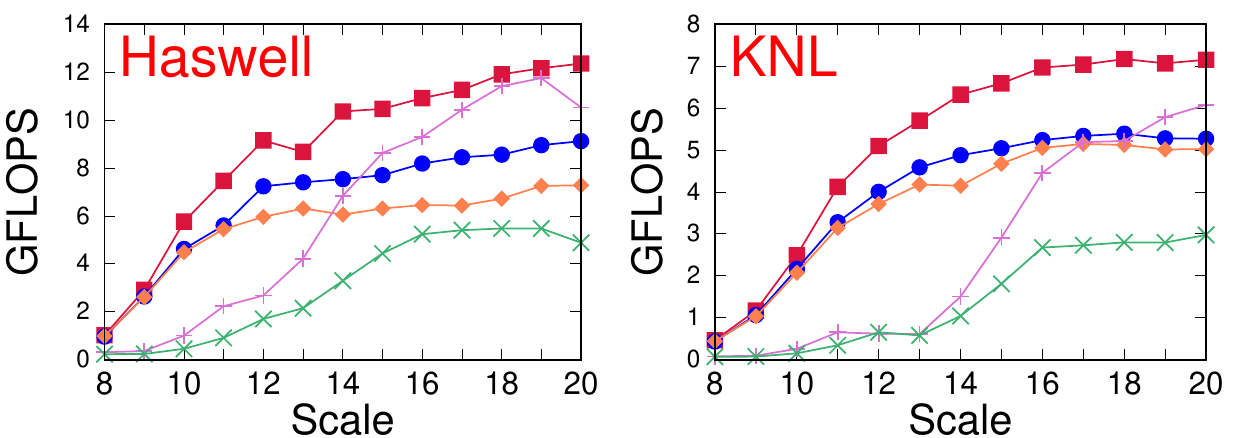}
	\end{subfigure}
	\caption{Triangle Counting: varying R-MAT scale.}
%	\vspace{-.1in}
	\label{fig:tricnt-rmat-flops}
\end{figure}

\noindent \textbf{Scaling with Input Size.}
Figure \ref{fig:tricnt-rmat-flops} shows the  performance on Haswell and KNL for R-MAT matrices with scale ranging from 8 to 20. 
MSA-1P obtains the highest GFLOPS rates on both KNL and Haswell.
Hash-1P and MCA-1P are slower than MSA-1P but they have similar trends.
\SSGB{} algorithms have bad performance for small inputs, however, as input size increases, SS:SAXPY gets closer to MSA-1P.

% \begin{figure*}
% 	\begin{subfigure}[b]{0.45\textwidth}
% 		\centering
% 		\includegraphics[width=\textwidth]{plots/tricnt-rmat-haswell-flops.pdf}
% 		\label{fig:tricnt-rmat-haswell-flops}
% 	\end{subfigure}
% 	\begin{subfigure}[b]{0.45\textwidth}
% 		\centering
% 		\includegraphics[width=\textwidth]{plots/tricnt-rmat-knl-flops.pdf}
% 		\label{fig:tricnt-rmat-knl-flops}
% 	\end{subfigure}
% 	\caption{Triangle counting - RMAT scaling on KNL and Haswel with edge factor 16}
% 	\vspace{-.1in}
% 	\label{fig:tricnt-rmat-flops}
% \end{figure*}

\begin{figure}
\begin{subfigure}[b]{0.45\textwidth}
	\centering
		\includegraphics[width=0.80\textwidth]{plots/tricnt-legend.png}	
	\end{subfigure}
	\begin{subfigure}[b]{0.45\textwidth}
		\centering
		\includegraphics[width=\textwidth]{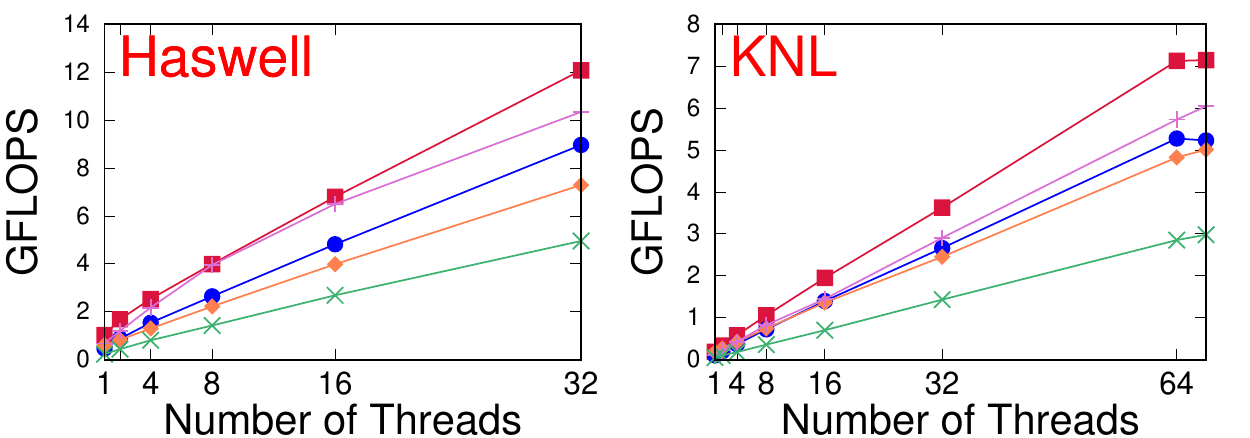}
	\end{subfigure}
	
	\caption{Triangle Counting: strong scaling (varying thread count) on R-MAT scale 20.}
%	\vspace{-.1in}
	\label{fig:tricnt-rmat-scaling-flops}
\end{figure}

\noindent
\textbf{Scaling with Thread Count.}
Figure~\ref{fig:tricnt-rmat-scaling-flops} shows the scalability analysis on Haswell and KNL for R-MAT matrix with  scale 20, on up to 32  threads on Haswell, and up to 68 threads on KNL, with all algorithms scaling well in all cases.

\subsection{$k$-truss}
For $k$-truss benchmark, we use $k=5$ and  report the sum of flops required to perform all Masked SpGEMM operations divided by total time required to execute them.
% K-Truss benchmark counts the number of k-trusses in a graph (in our case 5-trusses, so we refer to the bencmark as 5-truss in our figures).
% To find the k-trusses, we need multiple Masked SpGEMM kernel calls.

\begin{figure}
\begin{subfigure}[b]{0.45\textwidth}
		\centering
		\includegraphics[width=0.8\textwidth]{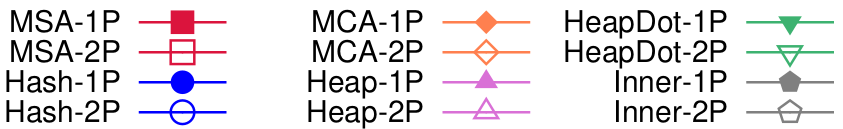}
	\end{subfigure}
	\begin{subfigure}[b]{0.5\textwidth}
		\centering
		\includegraphics[width=\textwidth]{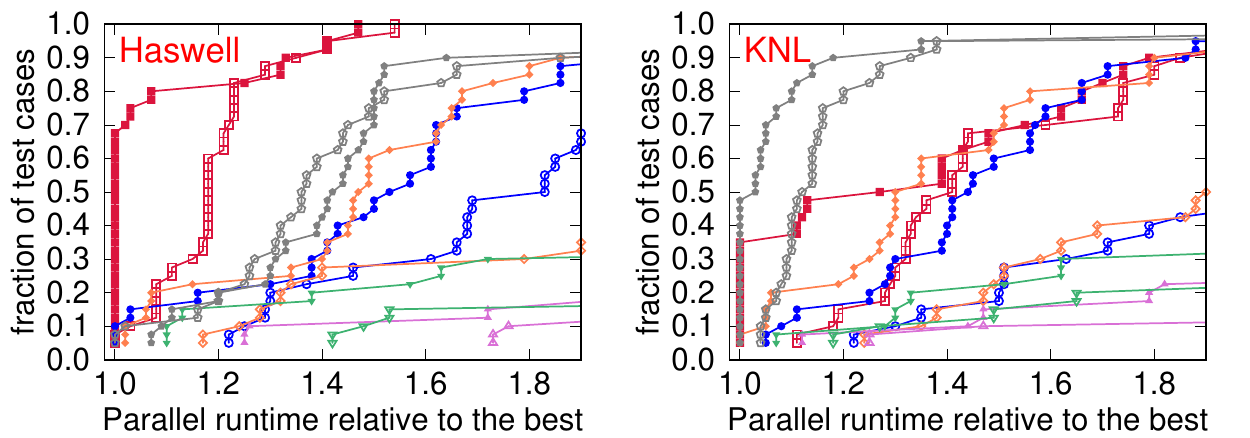}
	\end{subfigure}
	\caption{$k$-truss: Performance of the proposed schemes}
%	\vspace{-.1in}
	\label{fig:ktruss-knl26-1p2p}
\end{figure}

\noindent\textbf{Relative Performance of Masked SpGEMM Algorithms.}
Figure~\ref{fig:ktruss-knl26-1p2p} shows the performance of all our algorithms on all real graphs except wb-edu (excluded for its long running time).

MSA performs the best on Haswell while Inner performs fairly well on both, likely due to the mask getting sparser as $k$-truss  prunes the graph more with each iteration.
%
%\Oguz{Srdjan confirm this $\rightarrow$}
MSA's better performance on Haswell can be attributed to the existence of a large L3 cache (40 MB, whereas KNL has no L3 cache), hiding the cache misses due to large accumulator arrays in MSA to an extent.
The 1P schemes again perform better than 2P. Heap-based methods are noncompetitive, so we exclude them from our plots in the rest of this section.

% MSA-1P is the best performing algorithm on Haswell whereas Inner product is the best algorithm on KNL.
% They are followed by MSA and Hash.
% Heap and HeapDot were generally the slowest algorithms and they are excluded from the other plots.

\begin{figure}
\begin{subfigure}[b]{0.45\textwidth}
		\centering
		\includegraphics[width=0.8\textwidth]{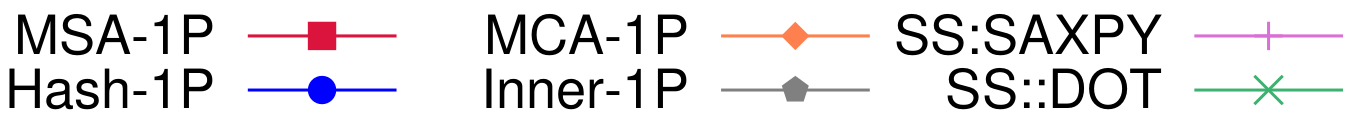}
	\end{subfigure}
	\begin{subfigure}[b]{0.5\textwidth}
		\centering
		\includegraphics[width=\textwidth]{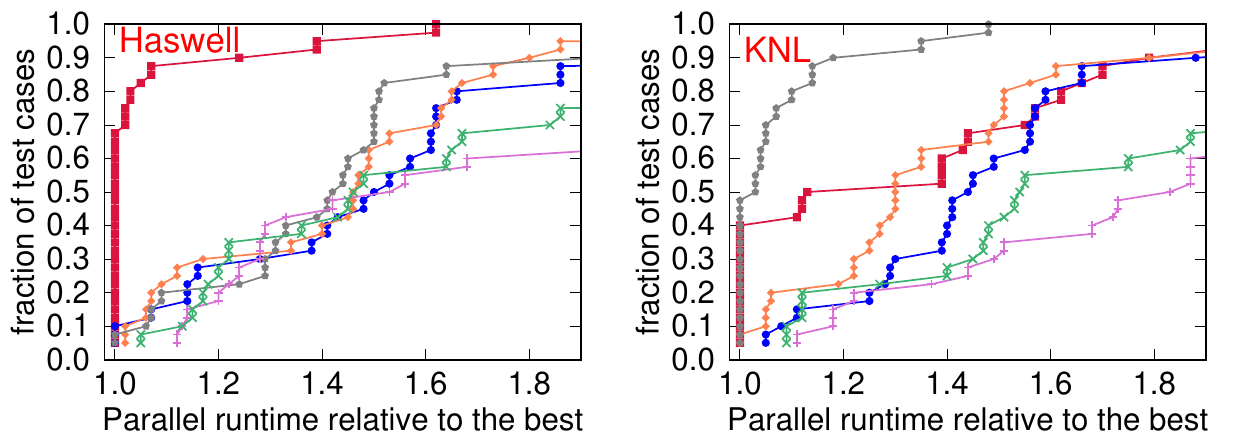}
	\end{subfigure}
	\caption{$k$-truss: our algorithms vs. \SSGB{}}
	\label{fig:ktruss-knl26-summary}
\end{figure}

Figure~\ref{fig:ktruss-knl26-summary} compares the performance of our four best performing algorithms against the \SSGB{} algorithms.
Our schemes MSA-1P and Inner-1P perform significantly better than \SSGB{} schemes on Haswell and KNL, respectively.

% MSA-1P is the best performing algorithm on Haswell and it is followed by all other algorithms.
% Inner product is the best performing algorithm on KNL, followed by MSA-1P, MCA-1P and Hash-1P.
% All our algorithms outperform \SSGB{} algorithms on KNL.

%\begin{figure}
%	\centering
%	\includegraphics[width=0.45\textwidth]{plots/ktruss-knl26-haswell-summary-perf-prof.pdf}
%	\caption{$k$-truss: our algorithms vs. \SSGB{}}
%	\vspace{-.1in}
%	\label{fig:ktruss-knl26-haswell-summary}
%\end{figure}
%
%\begin{figure}
%	\centering
%	\includegraphics[width=0.45\textwidth]{plots/ktruss-knl26-knl-summary-perf-prof.pdf}
%	\caption{$k$-truss: our algorithms vs. \SSGB{}}
%	\vspace{-.1in}
%	\label{fig:ktruss-knl26-knl-summary}
%\end{figure}

\begin{figure}
    \begin{subfigure}[b]{0.35\textwidth}
		\centering
		\includegraphics[width=0.75\textwidth]{plots/ktruss-legend.png}
	\end{subfigure}
	\begin{subfigure}[b]{0.45\textwidth}
		\centering
		\includegraphics[width=\textwidth]{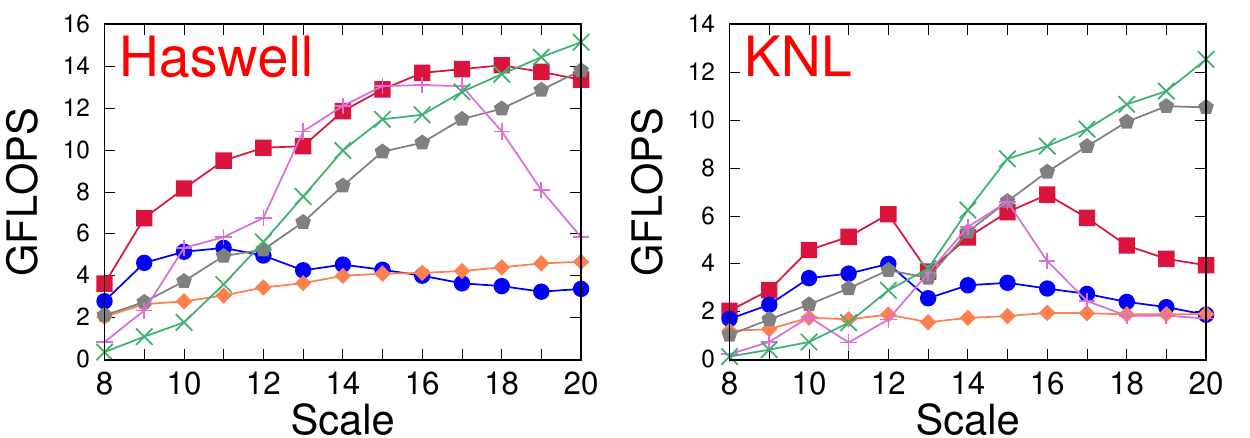}
	\end{subfigure}
	\caption{$k$-truss: varying R-MAT scale.}
%	\vspace{-.1in}
	\label{fig:ktruss-5-rmat-flops}
\end{figure}

\noindent
\textbf{Scaling with Input Size.}
Figure \ref{fig:ktruss-5-rmat-flops} shows the algorithm performance on Haswell and KNL for R-MAT matrices with scale ranging from 8 to 20. 
Inner and SS:DOT increase their GFLOPS rate well with increasing matrix scale, while MSA-1P does this only on Haswell.
The pull-based algorithms seem to attain better GFLOPS rates in the $k$-truss benchmark.
This benchmark shows that the algorithms that are deemed inefficient for plain SpGEMM can attain quite good performance when mask becomes part of the multiplication and can lead to highest GFLOPS rates.

% On Haswell, MSA-1P is the fastest algorithm when the scale is 18 or lower.
% \SSGB{} DOT and Inner product overtake MSA-1P at scale 18 and 19, respectively.
% Performance on KNL is similar.
% MSA-1P is the best algorithm when the scale is 12 or lower.
% When the scale is 13 or greater, \SSGB{} DOT is the fastest algorithm, followed by InnerProduct. 

\subsection{Betweenness Centrality}
% Betweenness centrality benchmark  calculates a node influence in a graph, using a
% batch variant which simultaneously calculates influences for multiple nodes.
Betweenness Centrality consists of a forward and backward stage, 
%both performing Masked SpGEMM, 
and uses both a complemented and non-complemented Masked SpGEMM.
%
% The forward stage utilizes a complemented Masked SpGEMM while the 
% Betweenness centrality consists of a forward phase and a backward phase.
% In the forward phase we execute multi-source BFS which uses Masked SpGEMM with complemented mask, and the backward phase uses plain Masked SpGEMM.
For this benchmark, we use TEPS~\cite{bader2006hpcs}, which is $\textit{batch\_size} \times \textit{num\_edges} / \textit{total\_time}$ as performance metric.
We use a batch size of 512.
%(i.e., the number of source vertices), which are chosen at random.

\begin{figure}
	\centering
	\includegraphics[width=0.4\textwidth]{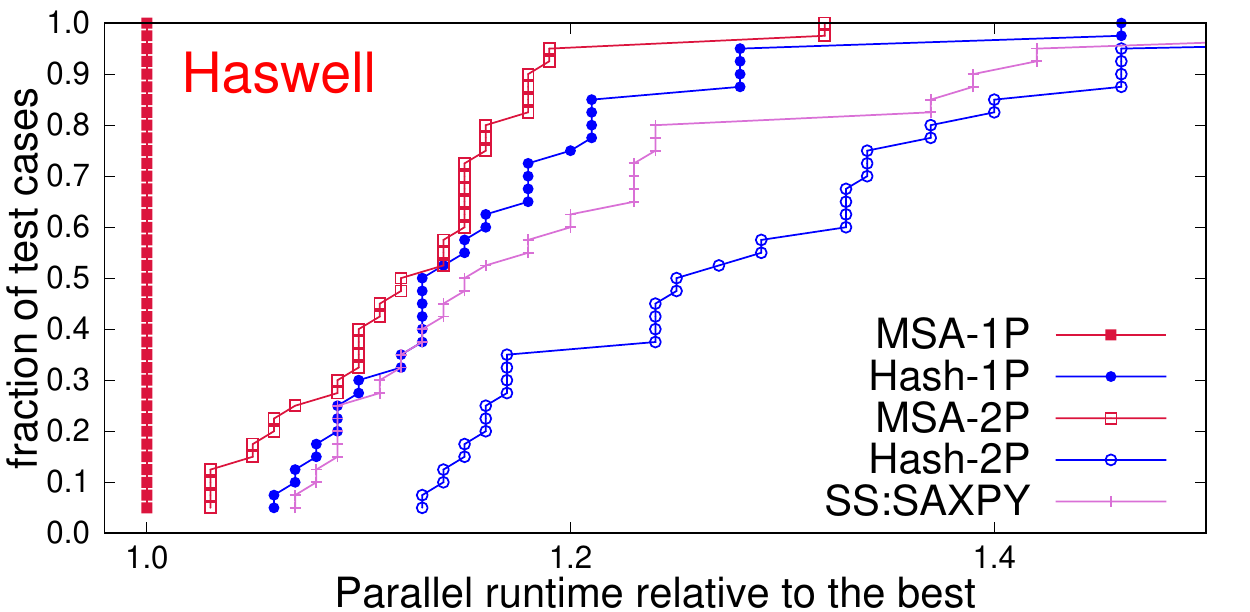}
	\caption{Betweenness Centrality: our schemes vs. \SSGB{}.}
%	\vspace{-.1in}
	\label{fig:bc-512-knl26-knl-summary}
\end{figure}

\noindent\textbf{Relative Performance of Masked SpGEMM Algorithms.}
Figure~\ref{fig:bc-512-knl26-knl-summary} shows the performance of all our algorithms on all real graphs except cage15, delaunay\_n24, and wb-edu (excluded for their long running time).
We benchmarked the Masked SpGEMM in forward and backward stages separately, but the trends were similar so we only present the overall performance.
MCA is not included in these results because it does not support complemented Masked SpGEMM.
We excluded Heap, Inner, and SS:DOT since they were prohibitively slow.

In this benchmark, MSA-1P obtains the best performance in \emph{all} test instances.
1P schemes again outperform 2P.
%

% MSA-1P is the fastest algorithm for all inputs, it is followed MSA-2P, Hash-1P, \SSGB{} SAXPY, and Hash-2P. 
% We saw similar trends for forward and backward phases individually, but 
% excluded those plots due to space limitations.

\noindent
{\bf Scaling with Input Size.}
Figure~\ref{fig:tricnt-rmat-scaling-flops} shows the algorithm performance on Haswell and KNL for R-MAT matrices with scale ranging from 8 to 20. 
% Figure  shows the algorithm performance on Haswell for G500 matrices with scale from 8 to 20. 
%
The schemes based on push-based algorithms, i.e., MSA-1P, Hash-1P, and SS:SAXPY are able increase their MTEPS rate with increasing matrix scale.
The mask in Betwenness Centrality can get quite dense, so the poor cache utilization of SS:DOT becomes a very serious bottleneck. In addition, the matrix $\mB$ is transposed in the library before each Masked SpGEMM, increasing overhead.

\begin{figure}
    \begin{subfigure}[b]{0.3\textwidth}
		\centering
		\includegraphics[width=0.75\textwidth]{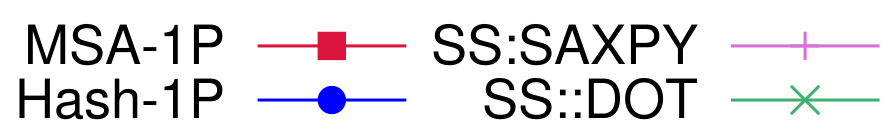}
	\end{subfigure}
	\begin{subfigure}[b]{0.45\textwidth}
		\centering
		\includegraphics[width=\textwidth]{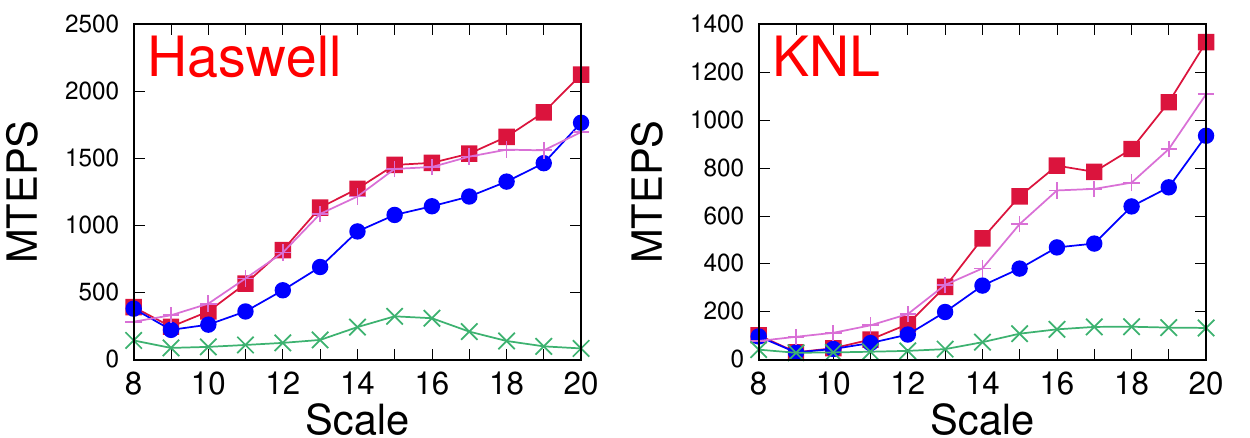}
	\end{subfigure}
	\caption{Betweenness Centrality: varying R-MAT scale.}
%	\vspace{-.1in}
	\label{fig:bc-rmat-scaling-flops}
\end{figure}

\section{Conclusions and Future Work}
\label{sec:conc}
In this paper, we presented four novel algorithms (with 10 total variations) for performing parallel masked sparse-sparse matrix multiplication.

We investigated Masked SpGEMM operation from various design and optimization standpoints to evaluate whether the challenges posed for plain SpGEMM still hold, and examined if some of the uncommon design choices can be reevaluated when mask is part of the multiplication.
We discovered that the mask and matrix density both have a critical effect on the performance of different design choices.
Surprisingly, we discovered that inner-product-based algorithm can be competitive in certain benchmarks with high mask sparsity on systems with small cache size.
We have shown that computing the Masked SpGEMM in a single phase usually performs better than approaches in which a symbolic multiplication is run prior to actual multiplication, in stark contrast with the conventions of plain SpGEMM computations where two-phase approaches are often more preferable.
We ran extensive experiments on two very different machines, with large sets of matrices, on several real-world benchmarks, and demonstrated that in almost all cases our methods significantly outperform the SuiteSparse:GraphBLAS~\cite{Davis2019} library, which is, to the best of our knowledge, the fastest Masked SpGEMM implementation in existence to-date.

As future work, we will investigate hybrid algorithms that can use different accumulators in the same Masked SpGEMM depending on the density of the mask and parts of matrices being processed, as well as exploiting fine-grain parallelism within single row processing.
% %
% We also would like to enhance our approaches for finer-granularity parallelism where threads can process parts of rows.
% %
% This is especially important for skewed graphs where a small number of vertices have a very high degree.

% We presented new multicore parallel algorithms for the masked SpGEMM problem. Our algorithms generally outperform the highly optimized SuiteSparse:GraphBLAS implementation for the same problem.
\begin{acks}
This work is supported by the Office of Science of the DOE under contract number DE-AC02-05CH11231. We used resources of the NERSC supported by the Office of Science of the DOE under Contract No. DE-AC02-05CH11231. 
\end{acks}
%\clearpage
\bibliographystyle{plain}
\bibliography{references}
\end{document}